\documentclass[fleqn,usenatbib]{mnras}

\usepackage[T1]{fontenc}
\usepackage{ae,aecompl}

\usepackage{graphicx}	
\usepackage{amsmath}	
\usepackage{amssymb}	

\newcommand{\rev}[1]{\textcolor{black}{#1}}
\newcommand{\revtwo}[1]{\textcolor{black}{#1}}
\newcommand{\revthree}[1]{\textcolor{black}{#1}}

\usepackage{newtxtext,newtxmath}

\title[New Satellites in GAMA Groups]{Galaxy And Mass Assembly: The xSAGA Galaxy Complement in Nearby Galaxy Groups}

\author[B.W. Holwerda et al.]{
B.W. Holwerda$^{1}$,\thanks{E-mail: benne.holwerda@louisville.edu}
S. Phillipps$^{2}$,
S. Weerasooriya$^{3}$,  
M. S. Bovill$^{3}$, 
S. Brough$^{4}$,
\and
M. J. I. Brown$^{5}$,
C. Robertson$^{1}$, 
K. Cook$^{1}$\\
$^{1}$ Department of Physics and Astronomy, University of Louisville, Natural Science Building 102, 40292 KY Louisville, USA\\
$^{2}$ Astrophysics Group, School of Physics, University of Bristol, Tyndall Avenue, Bristol BS8 1TL, UK. \\
$^{3}$ Department of Physics and Astronomy, Texas Christian University, Fort Worth, TX 76109, USA\\
$^{4}$ School of Physics, University of New South Wales, NSW 2052, Australia\\
$^{5}$ School of Physics and Astronomy, Monash Centre for Astrophysics (MoCA), 
Monash University, Clayton, Victoria 3800, Australia
}

\date{Accepted XXX. Received YYY; in original form ZZZ}

\pubyear{2015}

\begin{document}
\label{firstpage}
\pagerange{\pageref{firstpage}--\pageref{lastpage}}
\maketitle

\begin{abstract}
Groups of galaxies are the intermediate density environment in which much of the evolution of galaxies is thought to take place. In spectroscopic redshift surveys, one can identify these as close spatial-redshift associations. However, spectroscopic surveys will always be more limited in luminosity and completeness than imaging ones. 
Here we combine the Galaxy And Mass Assembly group catalogue with the extended Satellites Around Galactic Analogues (xSAGA) catalogue of Machine-Learning identified low-redshift satellite galaxies. 
We find 1825 xSAGA galaxies within the bounds of the GAMA equatorial fields ($m_r < 21$), 1562 of which could have a counterpart in the GAMA spectroscopic catalogue ($m_r < 19.8$). Of these, 1326 do have a GAMA counterpart with 974  below z=0.03 (true positives) and 352 above (false positives). 
By crosscorrelating the GAMA group catalogue with the xSAGA catalogue, we can extend and characterize the satellite content of GAMA galaxy groups. We find that most groups have $<$5 xSAGA galaxies associated with them but richer groups may have more. Each additional xSAGA galaxy contributes only a small fraction of the group’s total stellar mass ($<<$10\%).
Selecting GAMA groups that resemble the Milky Way halo, with a few ($<4$) bright galaxies, we find xSAGA can add a magnitude fainter sources to a group and that the Local Group does not stand out in the number of bright satellites. We explore the quiescent fraction of xSAGA galaxies in GAMA groups and find a good agreement with the literature. 
\end{abstract}




\section{Introduction}

Galaxies are social; most of them are found in groups of varying sizes \citep{Eke04,Robotham11} \rev{and truly isolated individual galaxies are very rare} \citep{Alpaslan14,Alpaslan15}. The most extensive and massive systems are relatively easily identified in even a sparsely sampled redshift survey. However, small and compact groups are easily missed in such single-pass surveys, as fibre collisions prevent more than one group member making it into the catalogue \citep{Robotham10}.

For groups of galaxies, \rev{an accurate} redshift \rev{and thus distance} and multi-band photometry are essential to measure the stellar content and \rev{to infer} to first order the group's halo mass \revthree{from the enseble's kinematics}. 
However, {\em dynamical} masses need a measure of the velocity dispersion of such a group and {\rev{hence one needs as many low-mass satellite galaxies (i.e. test masses in a dynamical system)} as practical. Small groups are the most common but their dynamical mass (and hence dark matter content) are the least constrained \revthree{from kinematics}. The best studied example of such a group is the Local Group, the complex of the Milky Way, Andromeda, M33 and their subsidiary satellites. The central question is however how representative this Local Group is \citep[e.g., in terms of galaxy occupation statistics,][]{Boylan-Kolchin11, Lovell11, Weisz11c,Robotham14, Boylan-Kolchin16}. The makeup of the Local Group is still very much in development with lower mass and low surface brightness members still being discovered in deep imaging surveys. \citep{Bechtol15, Drlica-Wagner15,Drlica-Wagner20, Kim15c, Koposov15, Homma19}. See for a review \cite{Simon19a} and for an overview of deep imaging results \cite{Wang21i}.

Here, our aim is to provide a context for the Local Group through statistics of the dynamical masses including a census of satellites, similar to the Satellites Around Galactic Analogues (SAGA) project \citep{Geha17} but starting from the Galaxy and Mass Assembly (GAMA) group catalogue \citep{Robotham11} \revthree{and supplementing it with Machine Learning identified satellites from the eXtended Satellites Around Galactic Analogues \citep[xSAGA,][]{Wu22d}. We aim to verify the ML prediction in the xSAGA catalogue with a GAMA redshift where possible and to evaluate the properties of added xSAGA satellites to GAMA groups. }

Cosmological $\Lambda$CDM simulations that attempt to reproduce galaxy groups (specifically the Local Group) struggle with the satellite distribution \citep[e.g., satellite alignment][]{Hammer13, Pawlowski13} 
but also the mass of the Magellanic Clouds \citep{Benson02, Koposov09, Okamoto10}. The simulations that match the Local Group's properties require a very quiescent environment and a ``quiet'', i.e. few mergers, assembly history of Andromeda and the Milky Way \citep{Kravtsov02, Klypin02, Gottloeber10, Forero-Romero11}. \rev{Reproducing the Local Group's history from observed satellite populations is an active area of research and a touchstone test for cosmological simulations \citep{Collins12,Collins14a,Collins16,Creasey15,Fattahi13,Fattahi16,Sawala16,Starkenburg17,Elias18,Garrison-Kimmel14,Garrison-Kimmel19a,Garrison-Kimmel19b,Digby19,Libeskind11,Libeskind20}.}
This all points to the possibility that the best-studied galaxy group with three massive members is --in fact-- very unusual and not representative of the group environment in our Universe in which most galaxies reside.
On the other hand, a sustained search for dwarf galaxy systems \citep{Geha12,Geha17,Mao21,Wang21i,Carlsten21,Carlsten22a,Carlsten22b} is showing that the Milky Way and M31 systems of satellites are not that unusual in their properties (e.g fraction of quenched satellites).

Recently, the GAMA survey \citep[Galaxy And Mass Assembly][]{Driver11} has made great inroads into identifying smaller groups of galaxies reliably \citep{Robotham11, Robotham12, Robotham13, Robotham14}. The impressive completeness of GAMA, 97\% of all $r<19.8$ magnitude sources in the field have spectroscopic redshifts \citep{Liske15,Driver22}, is enough to identify the lower-mass galaxy groups reliably. For example, one can look for Local Group analogues, with up to three or four bright galaxies and many smaller ones. However,  membership for the fainter galaxies remains uncertain because these typically do not have spectroscopic redshift measurements, even in the GAMA survey. Photometric redshift values are too uncertain: these cannot distinguish between distant background galaxies and faint group members. 

A second development is the recent extension of the SAGA survey \citep{Geha17} by \cite{Wu22d} using machine learning classifications of faint galaxies to successfully classify these as either distant background or belonging to a single halo surrounding a Milky Way analogue (xSAGA). Here, we start with the GAMA group catalogue ({\sc gamagoups}v10 from GAMA DR4) and cross-correlate with the xSAGA catalogue to compare the number, size, colour and surface brightness of xSAGA galaxies as a function of GAMA group properties. 

This paper is organized as follows: 
section \ref{s:gamagroups} describes the GAMA group catalogue, 
section \ref{s:xsagacat} describes the xSAGA catalogue and the comparison with GAMA group positions, 
section \ref{s:results} describes the results of the comparison between the xSAGA and GAMA group catalogues,
section \ref{s:conclusions}  discusses these results and places them in larger context with the aid of some group catalogues, and summarises our results and conclusions.
We assume a $\Lambda$CDM cosmology with the Planck18 cosmological parameters \citep[$H_0=67.4$ km/s/Mpc, $\Omega_m=0.315$,][]{Planck-Collaboration18b}

\begin{figure*}
    \centering
    \includegraphics[width=\textwidth]{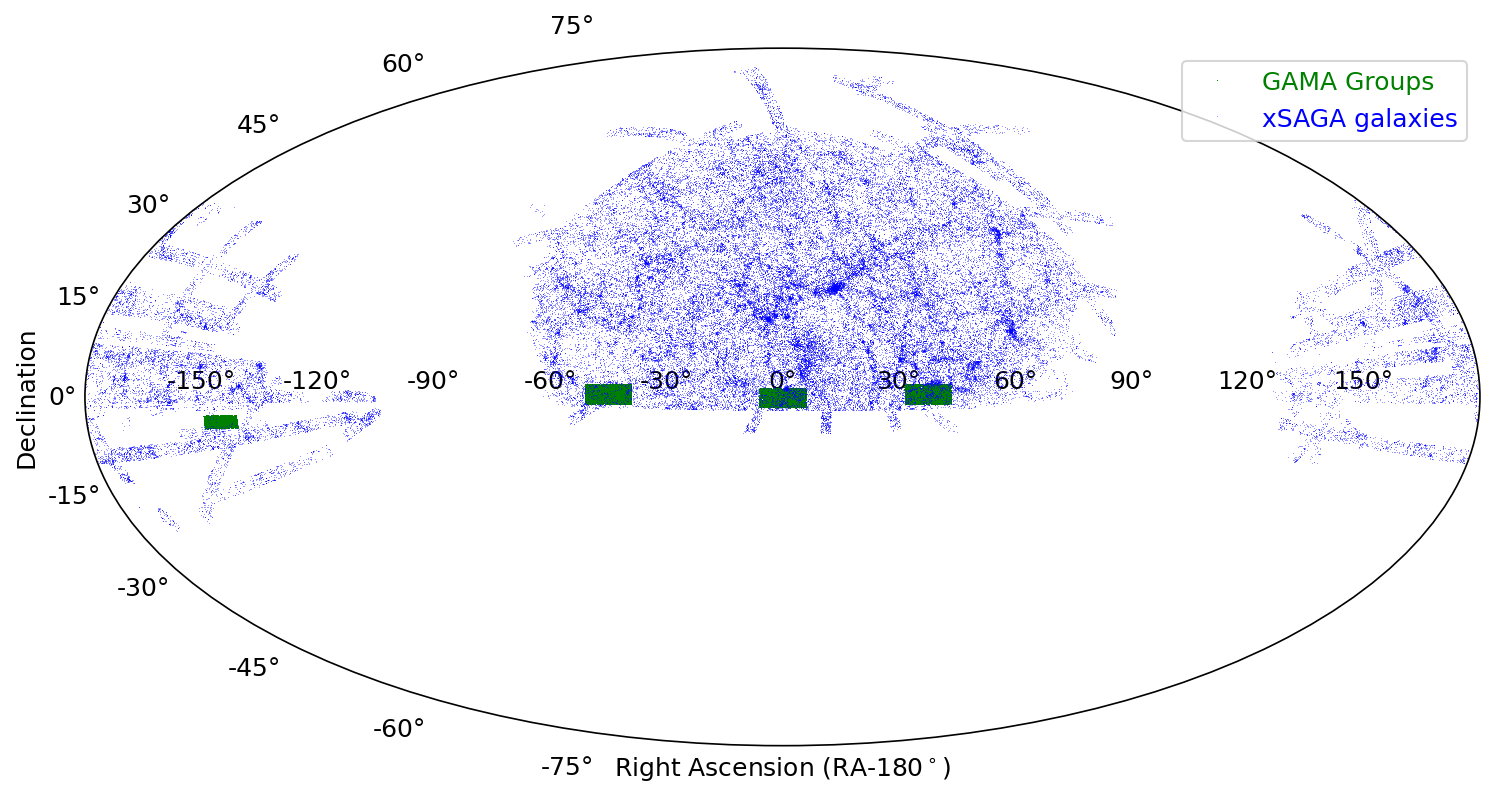}
    \caption{The position of the GAMA groups (green) and the xSAGA galaxies (blue). \protect\rev{The GAMA equatorial fields (green rectangles in the middle) are complete and the best studied ones. The GAMA group catalog includes a fourth field but this is not as  complete in spectroscopic redshifts nor is it fully covered by xSAGA. For simplicity, we only analyze the equatorial fields here.} }
    \label{f:ra-dec}
\end{figure*}

\section{GAMA Group Selection}
\label{s:gamagroups}

\rev{
Our starting point is the Galaxy And Mass Assembly (GAMA) survey \citep{Driver09, Liske15}. GAMA is a highly complete ($>$97\% to $r < 19.8$ mag) spectroscopic and multi-wavelength imaging survey conducted with the intent to investigate large-scale structure (LSS) in the local Universe ($z < 0.6)$ on kpc to Mpc scales \citep{Driver09, Driver11, Baldry18, Driver22}. The survey now consists of five survey regions, three of which are equatorial, covering a total of nearly 250,000 galaxies. Additional photometric data was collected on each galaxy in 20+ bands at multiple wavelengths \citep{Liske15, Driver16, Baldry18, Driver22}. }
This specific study uses GAMA Data Release 4, detailed in \citet{Driver22}. This highly complete redshift catalogue is ideal to identify smaller groupings of galaxies and local environments  \citep{Brough11a,Robotham11,Robotham14}. \rev{\cite{Robotham11} constructed the galaxy group and pair catalogue using a friends-of-friends algorithm. }

From the \cite{Robotham11} catalogue, updated for GAMA DR4 groups (gamagroupv10), we select all galaxy groups with a Friends-Of-Friends (FOF) redshift centre below z=0.03. This results in a selection of 272 groups. Figure \ref{f:ra-dec} shows their position on the sky. Figure \ref{f:Ngal-in-group} shows the distribution of the number of GAMA galaxies in these groups. The majority of groups has 2-3 massive members in them, making them very similar to the Local Group. 
\begin{figure}
    \centering
    \includegraphics[width=0.5\textwidth]{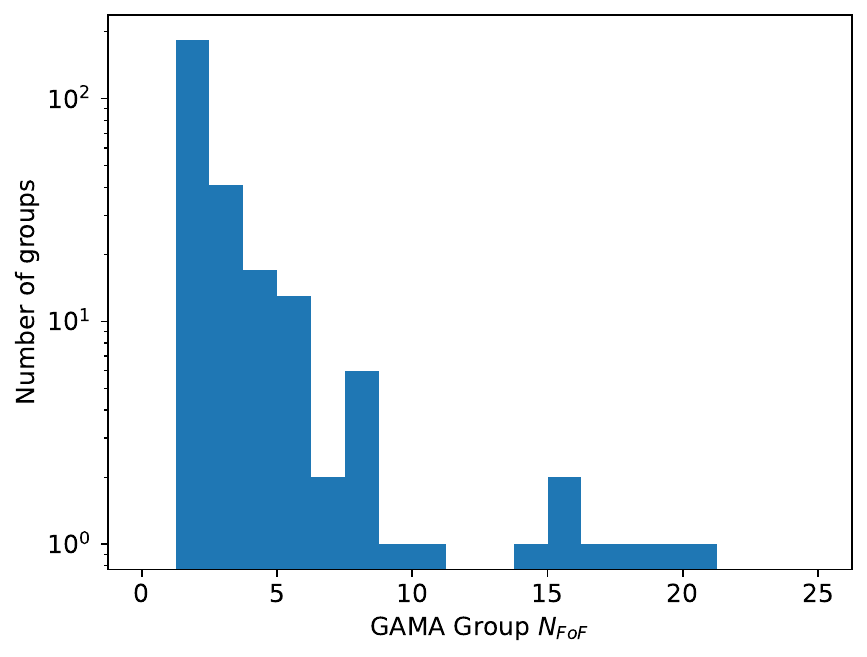}
    \caption{The number of galaxies in GAMA groups below $z<0.03$. The majority are comprised of a few galaxies.}
    \label{f:Ngal-in-group}
\end{figure}

\section{xSAGA dwarf catalogue}
\label{s:xsagacat}

To quantify if a Milky Way analogue galaxy \revthree{or a Local Group analogue} has a similar retinue of satellite galaxies, one needs highly complete spectroscopic catalogues around every potential Milky Way+Andromeda grouping. In the local volume this is the target for the ELVES survey \citep[][]{Carlsten21,Carlsten22a,Carlsten22b}. For much larger volume searches, incompleteness plagues the accurate characterization of the retinue of satellite galaxies. 
Thus a highly complete satellite catalogue has been the science objective of the Satellites Around Galactic analogues \citep[SAGA,][]{Geha17} survey. However, this is observationally extremely expensive as it requires high-resolution spectroscopic redshift confirmation of intrinsically faint galaxies. 

\cite{Wu22d} have produced a catalogue of dwarf galaxies based on the Dark Energy Spectroscopic Survey (DESI) imaging \citep{Dey19} using a machine learning technique. Their aim is to vastly expand the number of high-probability dwarf systems surrounding Local Group analogues. \cite{Wu22d} used the spectroscopic redshift catalogues of the SAGA Survey to identify a training data set, and then optimized a convolutional neural network (CNN) to distinguish $z < 0.03$ galaxies from more-distant objects using image cutouts from the DESI Legacy Imaging Surveys \revthree{as input}. 
\cite{Wu22d} identify a sample of over 100,000 CNN-selected low redshift galaxies with CNN probabilities greater than 50\% to be located at low redshift ($P_\mathrm{z<0.03} > 50$\%, Figure \ref{f:Mgroup:Pcnn}). This is the extended SAGA catalogue (xSAGA). 
For xSAGA, \cite{Wu22d} claim to be complete to $M_r < -15$ mag and select objects with $m_r < 21$ mag. This is almost 1.5 magnitudes deeper than the GAMA cut-off \citep[][]{Liske15, Driver22}.  
This is the xSAGA catalogue against which we compare the low-redshift GAMA group positions and GAMA membership. GAMA was not used in the training of xSAGA and provides a completely independent confirmation of the CNN predictions by \cite{Wu22d}. \revthree{Redshift verification with GAMA spectroscopy of the Machine Learning low redshift prediction is a first goal for this work.}
\begin{figure}
    \centering
    \includegraphics[width=0.5\textwidth]{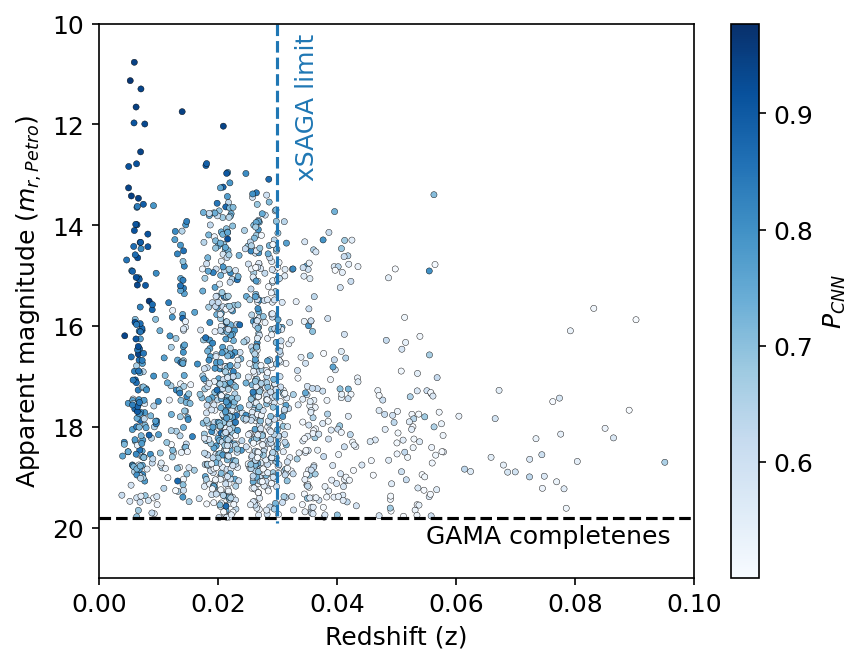}
    \caption{The redshift and r-band Petrosian apparent magnitude of xSAGA galaxies identified in \rev{the GAMA equatorial fields' footprint with GAMA catalogue counterparts}. The colour is indicative of the CNN confidence in the $z<0.03$ identification. Some 26.5\% are above the redshift limit and we note the majority of these are lower confidence. A more stringent cut in CNN confidence level would result in a cleaner but less complete sample.  }
    \label{f:xsaga:gama}
\end{figure}

\subsection{xSAGA Galaxies in GAMA }

First, we examine how many xSAGA galaxies are in the equatorial survey area of GAMA (\revthree{typically designated } as the G09, G12 and G15 fields). \rev{Table \ref{t:detections} summarizes these numbers.} We find that 1825 xSAGA galaxies are within these three fields RA and DEC limits. 
Secondly, we compared the positions of xSAGA galaxies against those of the GAMA catalogue (G3CGalv10). 
Of the xSAGA galaxies in the equatorial fields, 1586 are bright enough to potentially be selected for GAMA \citep[$m_r = 19.8$, 90\% \revtwo{complete,}][]{Driver22}. Of these possible GAMA members, 1326 have a counterpart in GAMA within two arcsecond radius,\revthree{the GAMA fibre aperture used on the Angle-Australian Telescope AAOmega instrument}, 260 do not (84\% complete). The 6\% difference \rev{between the reported GAMA completeness and the one in xSAGA/GAMA overlap} may be due to rejection \revthree{of some xSAGA sources} by their small angular size \revthree{in the original GAMA target selection}. 

Figure \ref{f:xsaga:gama} shows the distribution of spectroscopic redshift and apparent magnitude for the xSAGA overlap with GAMA. The CNN probability assigned to each xSAGA galaxy is \revthree{denoted by} the colour for each point. Of the overlap of xSAGA and GAMA, 974 are below $z=0.03$ according to the GAMA spectroscopic redshift (true positives) and 352 are not (false positive). 
The false positive rate in the xSAGA/GAMA overlap is 26.5\%. The expectations from \cite{Wu22d} were for 71\% precision (purity \revthree{or true positive rate}) with a trailing of redshift values above z=0.03. \cite{Wu22d} show in their Figure 3, how the CNN's precision declines below $m_r = 19$ mag. \revthree{Our true positive and false positive rates are consistent with the predicted true positive rate and the redshift behavior and worsening performance at lower luminosities predicted by \cite{Wu22d}}.  

It is not possible to estimate the true negative ($T_N$) and false negative ($F_N$) rates as we only have the xSAGA selection, not their initial candidates. The precision\footnote{Precision is defined as: $\rm  P = \left({\rm T_P \over T_P + F_P}\right)$, $T_P$ is the number of true positives and $F_P$ is the number of false positives.} is 974/1326 (73\%) but for \rev{other typical metrics of machine learning}, recall\footnote{Recall is defined as $\rm R=\left({\rm T_P \over T_P + F_N}\right)$, where $T_P$ is the number of true positives and $F_N$ is the number of false negatives.} and the compound F1 metric\footnote{The F1 metric is a combination of precision (P) and recall (R): $\rm F1= 2 \times \left({P \times R \over P + R} \right)$} we would need to know the \rev{true and false} negative rates as well. 
\revthree{The three metrics are often take together as any machine learning algorithm is a trade-off between precise predictions and a complete selection from a large sample (recall). A precision metric over 70\% is perfectly workable in a statistical sense but individual sources and thus group membership have a substantial chance still of false positives. }

\begin{table}
    \centering
    \begin{tabular}{ll}
Sample                                  & number of galaxies \\
\hline
\hline
xSAGA within 3 GAMA equatorial fields: 	& 1825 \\
xSAGA within GAMA detection envelope:   & 1586 \\
(in equatorial fields, $m_r < 19.8$)    & \\
xSAGA within 3 GAMA equatorial fields   & \\
not found in GAMA: 		                & 260 \\
xSAGA found in GAMA: 	                & 1326 \\
(in equatorial fields, $m_r < 19.8$)    & \\
xSAGA in GAMA below z=0.03              & 974 \\
xSAGA in GAMA above z=0.03              & 352 \\
\hline
    \end{tabular}
    \caption{The number of xSAGA galaxies in the equatorial fields, within the GAMA detection envelope, those found and not found in GAMA within the bounds of the equatorial fields and of those xSAGA galaxies with a counterpart in GAMA below and above the z=0.03 redshift limit that xSAGA was trained on. }
    \label{t:detections}
\end{table}

In the following, we mark the galaxies that are common to xSAGA and GAMA in black and the new xSAGA candidate group members in blue. 

\subsection{Group Cross Match}

We only consider GAMA groups that are \rev{at redshifts below} $z<0.03$ for cross match. We adopt the r-band luminosity weighted centre of the group system as the position of the group centre. Starting from the group centre on the sky (CenRA,CenDec), we counted the numbers of xSAGA satellites within both R100 and R50, the full and half-radii of the group, and inferred the distance to the nearest xSAGA galaxy for each group using {\sc match\_coordinates\_sky} in {\sc astropy.coordinates} \rev{to determine if they were within R100 or R50}. In the case of groups of only two members, the full radius is the projected distance between them and the half-radius is half of that. The R100 radius is below the 300 kpc cut-off \cite{Wu22d} use for group membership in training for almost all GAMA groups below z=0.03. \rev{There are 272 GAMA Groups in the group catalogue at redshifts below z=0.03. Of these, 190 have one or more xSAGA galaxies associated with them within the R100 radius.}
Each GAMA group has a mass estimate (MassA) and total luminosity estimate \citep[totRmag][]{Robotham11}. Figure \ref{f:Mgroup:Pcnn} shows the confidence level of the xSAGA CNN in the xSAGA galaxies associated with a GAMA group as a function of GAMA group mass. \revthree{At the lower mass end of the GAMA groups ($\rm log_{10}(M_{\rm group}/M_\odot) < 10$), the xSAGA satellites are rarer and all already in GAMA. 
Above this group mass, there is no dependence on CNN confidence in the xSAGA identification on either group mass or selection as a GAMA target. xSAGA selection does not bias against group masses above this mass nor whether they were targeted by GAMA.}

In the following, we compare the group characteristics, \rev{the number of galaxies in the group found via friends-of-friends, the dynamical mass of the group and the total sdss-r band luminosity ($N_\mathrm{fof}$, MassA, totRmag), to those characteristics observed in the xSAGA satellites.} \revtwo{MassA is used in the following figures as the mass of the GAMA group ($\rm log_{10}(M_{\rm group}/M_\odot)$) and totRmag as the measure of group luminosity} \revthree{for an estimate of the total stellar mass in section 4.3. }

\begin{figure}
    \centering
    \includegraphics[width=0.5\textwidth]{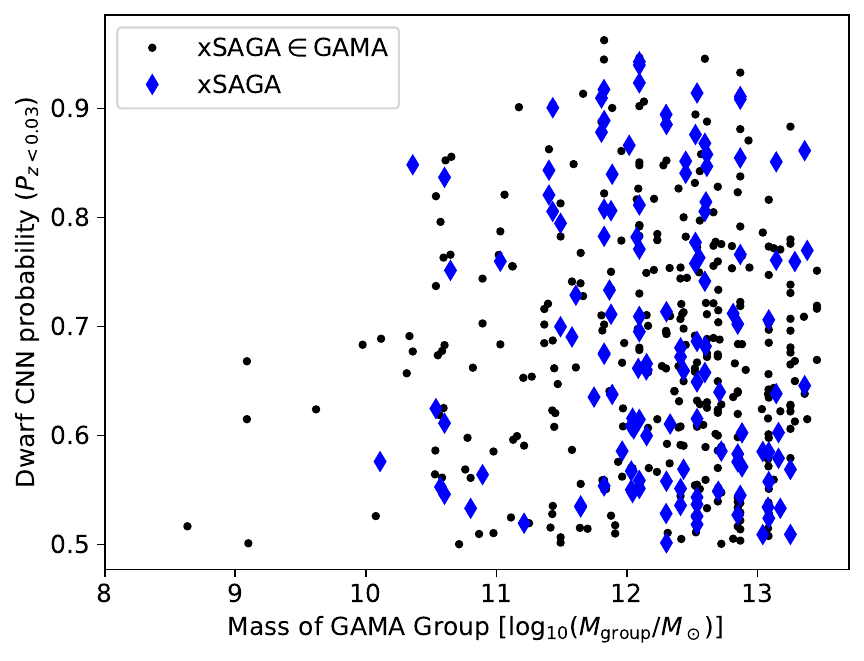}
    \caption{The CNN probability of an xSAGA galaxy being an $z<0.03$ galaxy as a function of its most likely associated GAMA group mass \revthree{\protect\citep[MassA from the ][catalogue]{Robotham11}}. These are all the xSAGA galaxies down to $m_r <21$ \rev{and within the R100 of a GAMA group at $z<0.03$}. The few xSAGA galaxies in low-mass GAMA groups have CNN confidence levels in the lower range of acceptability compared to higher mass ($\log_\mathrm{10}(M/M_\odot) > 10.5$) groups. }
    \label{f:Mgroup:Pcnn}
\end{figure}

\section{Results}
\label{s:results}

\revthree{The results of this exercise to expand the satellite tally in GAMA groups using xSAGA catalogue can be split into a few broad categories: }
what we learned about the GAMA groups, 
what we learned about the satellite galaxy population now associated with galaxy groupings, and how our Local Group satellite population compares to other, similar groups in broader surveys.}

\begin{figure*}
    \centering
    \includegraphics[width=0.32\textwidth]{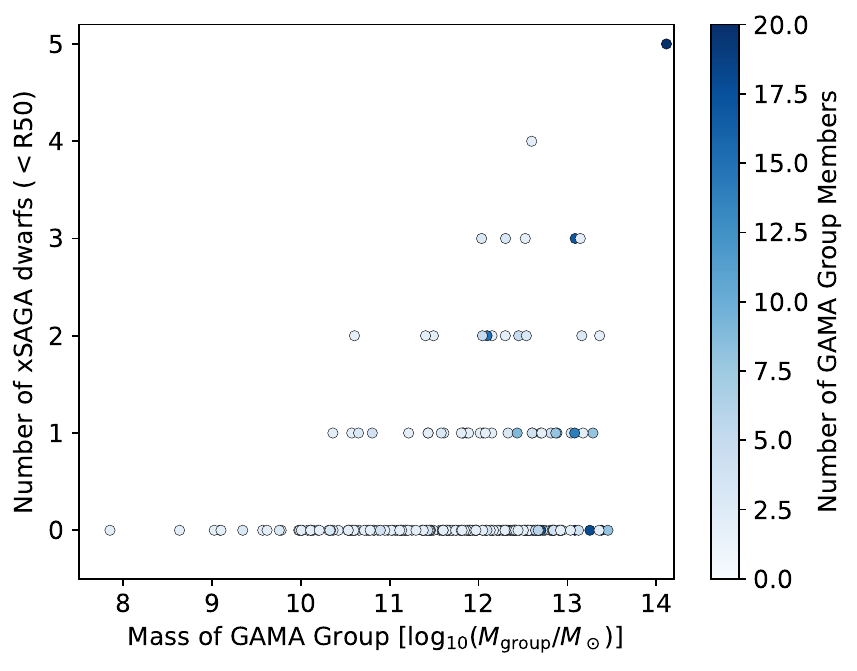}
    \includegraphics[width=0.32\textwidth]{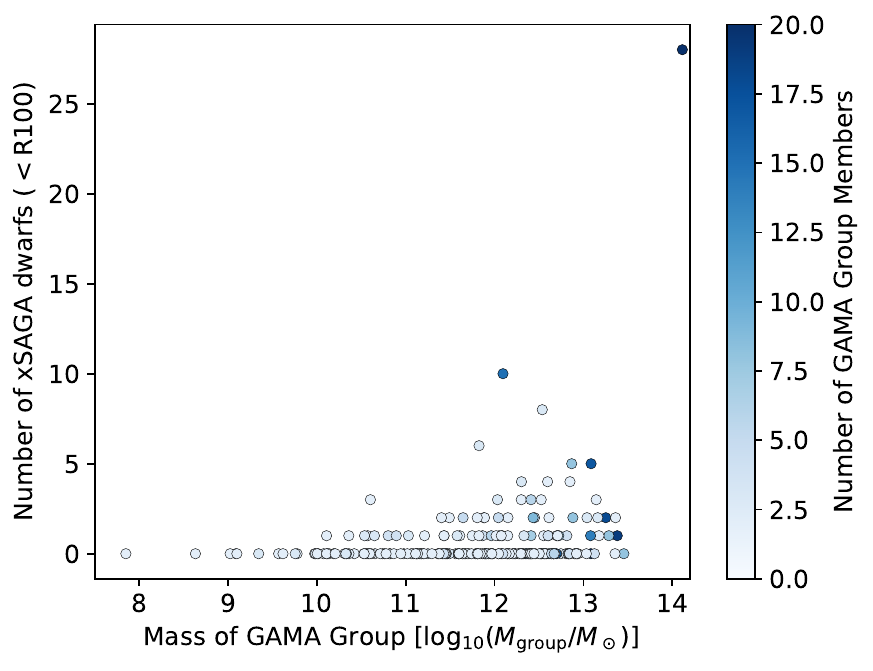}
    \includegraphics[width=0.32\textwidth]{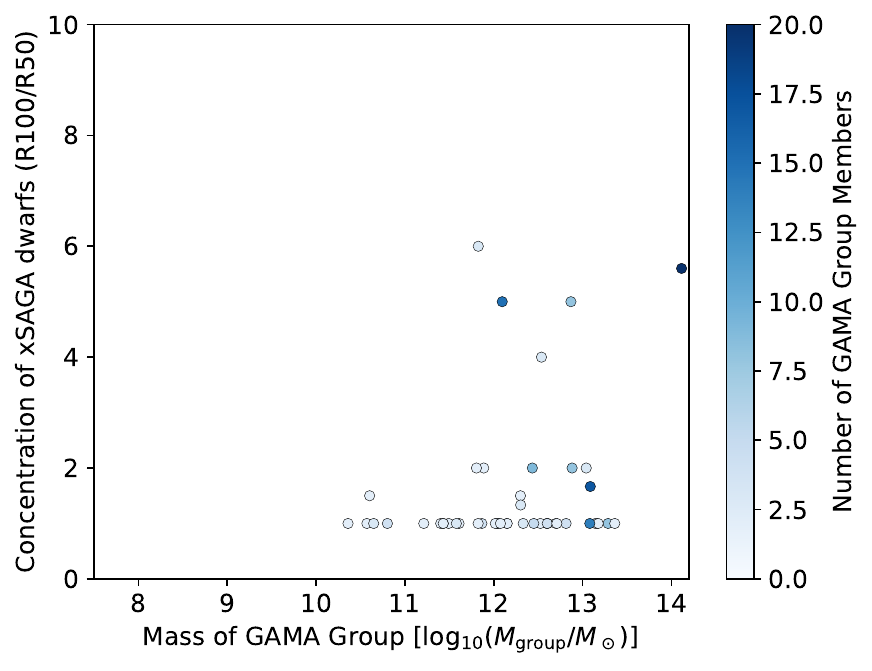}
    \caption{The number of new xSAGA galaxies found within the 50\% radius (left), 100\% radius (middle) of each GAMA group as a function of GAMA group mass \revthree{\protect\citep[MassA from the ][catalogue]{Robotham11}}. The concentration of new xSAGA candidate group members (right panel) is defined as the ratio of the number found within R100 divided by that within R50.  
    \revthree{Higher values of this ratio indicated a more spread-out satellite distribution within the R100 radius. Most groups concentrate their xSAGA retinue within R50, some have half within the R50 and the rest within the R100 radius. And only a few massive groups have the majority of their xSAGA satellites between R50 and R100.}
    \rev{The one group at $\log_\mathrm{10}(M_\mathrm{group}/M_\odot) \sim 14$ is both richer ($N_\mathrm{FoF} > 20$) and more enriched (27 xSAGA satellites within R100, middle panel). This could be considered a small cluster, rather than a sparse group.}}
    \label{f:Ndwarfs}
\end{figure*}

\subsection{Numbers of xSAGA satellites inside GAMA groups}

Here we look at the numbers of xSAGA galaxies that fall within a GAMA group's boundary, either the half or full fraction of galaxies radius (R50 or R100).  
Figure \ref{f:Ndwarfs} shows the number of new xSAGA galaxies within the half-size and full-size radius of the group as a function of mass. The number of friends-of-friends identified GAMA galaxies in the group is indicated with the colour-bar. The majority of GAMA groups do not have an xSAGA satellite associated with them within either the R50 or R100. More massive groups however often have a few xSAGA identified satellites. \rev{There is one outlier of 25 xSAGA galaxies in a massive and rich group of 20 GAMA members. }
Figure \ref{f:Ndwarfs} shows the concentration of xSAGA ($\notin$ GAMA) galaxies in groups, defined as \rev{follows:}
\begin{equation}
    C = {N_\mathrm{xSAGA} (< R100) \over N_\mathrm{xSAGA} (< R50) }
\end{equation}
i.e. the ratio of the number of xSAGA galaxies within R100 over R50. \rev{Lower values of the concentration index means the new xSAGA galaxies are concentrated more towards the inner parts of the group.} \revthree{Higher values of this ratio indicated a more spread-out satellite distribution but within the R100 radius. Most groups retain their xSAGA retinue within R50, some have half within the R50 radius and the other half within full R100 (concetration of 2). And only a few massive groups have the majority of their xSAGA satellites between the R50 and R100 radii.}
Lower mass groups are more compact with their xSAGA satellites compared to more massive ones. Most GAMA groups are N=2 but some have up to 20 members and additionally show a wider distribution of associated xSAGA galaxies. 
\rev{We note a single massive group ($ \log_\mathrm{10}(M_\mathrm{group}/M_\odot) \sim 14$, Group ID 200006) which is both richer ($N_\mathrm{FoF} = 67$) and more enriched (27 xSAGA satellites within R100). }
\revthree{Generally speaking, an increase in the spatial distribution of satellites can be expected with increased group mass. }

\begin{figure}
    \centering
    \includegraphics[width=0.5\textwidth]{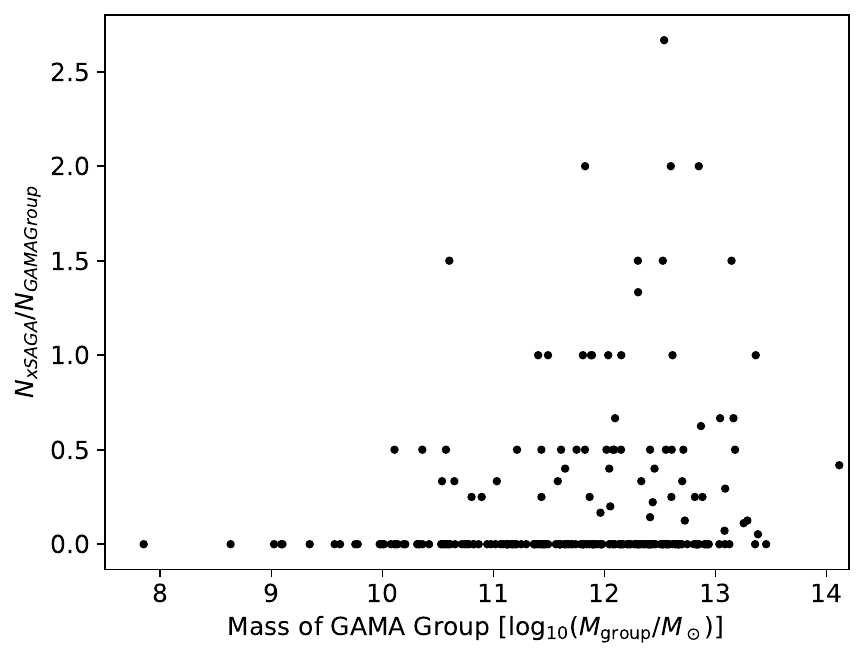}
    \caption{The ratio of the number of new xSAGA galaxies found within the 100\% radius of each GAMA group over the number of GAMA-defined group members as a function of GAMA group mass.}
    \label{f:dwarf:total}
\end{figure}

Figure \ref{f:dwarf:total} shows the ratio of associated xSAGA galaxies to the total GAMA identified group members. A number of the more massive groups gain up to 200\% more members. \revtwo{This illustrates that a reliable machine learning search in images for fainter group members, such as the xSAGA project \citep{Wu22d}, will result in a plethora of new sources and potential targets for redshift measurements to improve the kinematic mass estimate for any galaxy groups identified through a few bright members} \revthree{in a spectroscopic redshift survey. The economy of a machine learning approach to this observational problem is cleanly illustrated here.}

\subsection{xSAGA contribution to Group Stellar Mass}

How much additional stellar mass do these potential new members bring to the group? Figures \ref{f:Ndwarfs} and \ref{f:dwarf:total} suggest a substantial contribution in number of galaxies to the groups. Here, we convert the $g-r$ colour and SDSS-r apparent magnitude reported for the xSAGA galaxies to a stellar mass using the prescription for stellar mass for the $g-r$ colour from \cite{Zibetti09a}. We convert the total luminosity of the group in sdss-r band \citep[from totRmag in the GAMA group catalogue see][]{Robotham11} using the same colour for the sdss-r band M/L ratio. We thus arrive at a total group stellar mass and the stellar mass contribution by each xSAGA satellite. 

\begin{figure}
    \centering
    \includegraphics[width=0.5\textwidth]{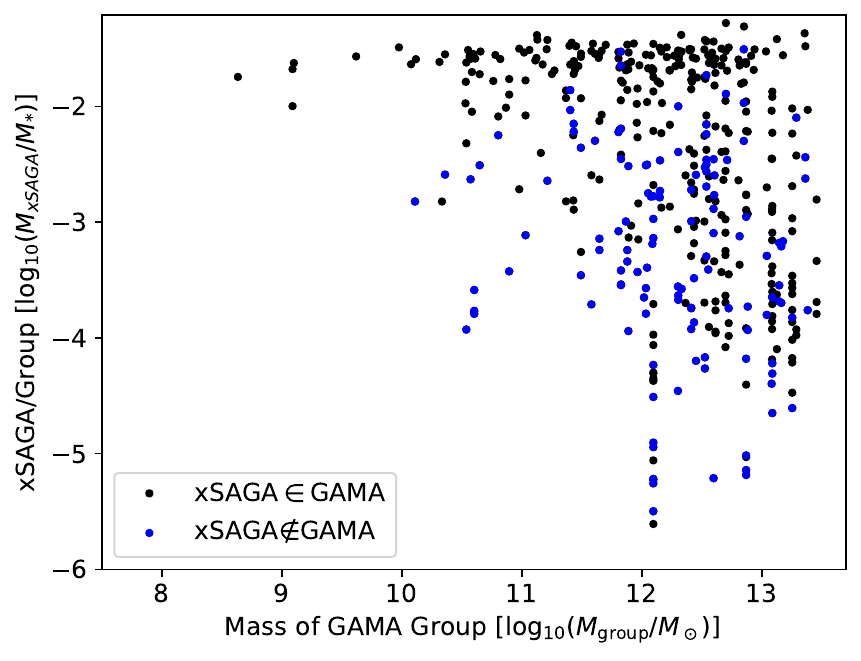}
    \caption{The fraction of the stellar mass that is contributed by xSAGA satellite galaxies to each GAMA group. Black dots are xSAGA galaxies with a GAMA counterpart, blue dots are without a GAMA catalogue counterpart.} 
    \label{f:Mgroup:xsagamassfrac}
\end{figure}

Figure \ref{f:Mgroup:xsagamassfrac} shows the fraction of stellar mass each xSAGA galaxy would have increased the group stellar mass by. \rev{The majority of the xSAGA galaxies already in GAMA contribute substantial fractions to the stellar mass of the group. Considering these are typically small groups, this is unsurprising. Most of the xSAGA galaxies not included in GAMA contribute a small fraction (less than 1\% of stellar mass) with some exceptions. The uncertainty on stellar mass contribution is substantial because we need to use a single colour to derive the mass-to-light ratio and the group's distance to compute the absolute magnitude, \rev{when a GAMA redshift is not available}. We note that xSAGA galaxies were associated with a group using projected radii. Even in the limited redshift range ($z<0.03$, 136 Mpc), there could still be false positives i.e. foreground xSAGA galaxies included in a distant group or a background xSAGA galaxy included in a foreground galaxy group count. } \revthree{Even with a precision of over 70\%, a number of the additional satellites will be spurious.}

\revthree{For context, the intragroup light (IGL) of a single GAMA group ($\log_\mathrm{10}(M_\mathrm{group}/M_\odot) \sim 13$) showed evidence for $<$36\% of the group's stellar mass in the diffuse IGL component \citep{Martinez-Lombilla23}. A few percent additional stellar mass fits the picture of slow present growth of the intragroup component if all these xSAGA satellites are eventually converted to the intragroup stellar population. }


\subsection{xSAGA Satellite Characteristics in GAMA Groups}

\begin{figure}
    \centering
    \includegraphics[width=0.5\textwidth]{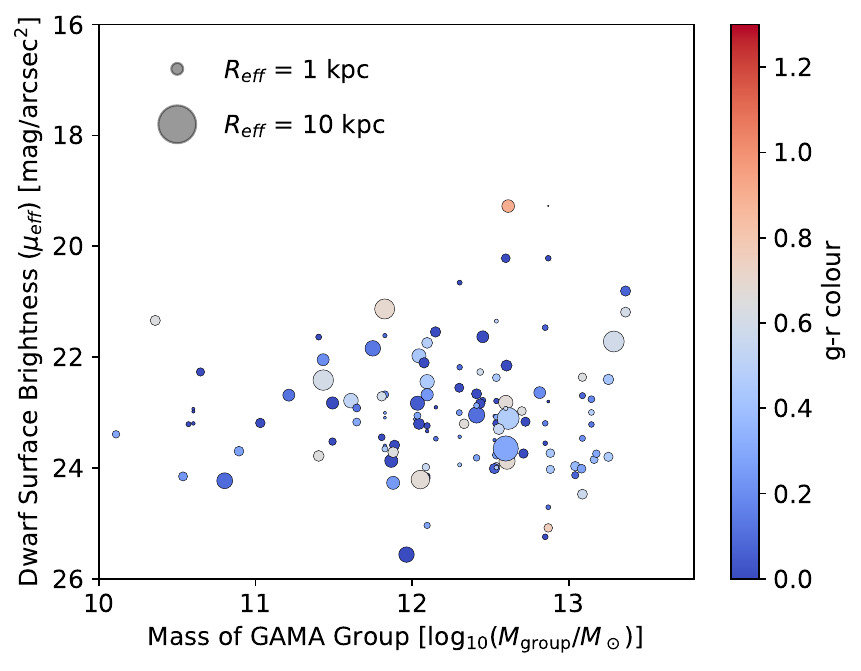}
    \caption{The surface brightness within the effective radius, $g-r$ colour, and effective radius of the xSAGA galaxies not found in GAMA, found within the R100 radius of the GAMA groups as a function of GAMA group mass. The majority of xSAGA galaxies are small, within a small range of surface brightness, and of a range of $g-r$ colour. The colour scale is to match the star-forming (blue) and quiescent (red) criterion of \protect\cite{Salim14a}. The majority of xSAGA galaxies are star-forming but there is a substantial quiescent population. }
    \label{f:Mgroup:mu-color-reff}
\end{figure}

In this section, we take a closer look at the characteristics of xSAGA satellites identified within the GAMA groups. 
Figure \ref{f:Mgroup:mu-color-reff} shows the effective surface brightness ($\mu_\mathrm{eff}$) of the xSAGA galaxies against the dynamical mass of the GAMA group \revthree{(MassA)}. \rev{The effective surface brightness is computed from the r-band apparent magnitude ($r_0$) and the effective radius by 
$m_\mathrm{r,eff} = r_0 + 2.5 \log[ 2\pi (R_\mathrm{r,eff}/\mathrm{arcsec})^2 ]$ \citep{Wu22d}. The effective radius ($R_\mathrm{r,eff}$) is the Petrosian r-band half-light radii, enclosing half the flux in the sdss-r band through a growth curve algorithm.}
The effective radii and effective surface brightness are from the DESI catalogue. xSAGA satellites' surface brightnesses and small effective radii ($r_\mathrm{eff} << 10 \ \rm kpc$) firmly put \revthree{the majority of} them in the lower surface brightness and dwarf galaxy regime.
\rev{The symbols in} Figure \ref{f:Mgroup:mu-color-reff} \rev{are colour-coded according to} the $g-r$ colour of the xSAGA galaxies. The colour scale is to highlight the separation between quiescent (red) and star-forming galaxies (blue) \revthree{as defined by \cite{Salim14}.} The majority of the xSAGA satellites is star-forming but quiescent galaxies are picked up as well in all group masses. 
\revthree{The selection of both red and blue galaxies by xSAGA shows that the initial colour selection and subsequent machine learning algorithm does not prohibitively bias against red galaxies. }

\subsection{The Quenched Fraction of xSAGA Satellites}

The dependence of star formation on group environment was a key science goal for the GAMA project \revthree{\citep[see][]{Driver09}.} 
\revthree{The general picture for \textit{isolated} dwarf galaxies is that these almost all star-forming \citep{Haines07,Geha12,Kawinwanichakij17} with only a small number quiescent isolated galaxies reported \citep[see][]{Monachesi14,Martinez-Delgado16,Garling20,Polzin21,Casey23}. 
In GAMA groups, }\cite{Treyer17a} examined the quenching of group central galaxies and found evidence for substantial quenching in satellites \citep[``conformity'', see also][]{Weinmann06,Wang12c,Kauffmann13a}. \revtwo{ \cite{Grootes17,Grootes18} found that gas cycling is similar to field spirals in the satellites of GAMA groups. They only find a lower specific star-formation rate in the massive satellites. \revthree{\cite{Sotillo-Ramos21} find no evidence of quenching in groups for lower mass galaxies.} \cite{Pearson21} find that galaxies in groups typically become larger with group halo mass and no evidence for dramatic changes in morphology with increasing group mass.} \cite{Davies19a} note that scatter around the star-formation and stellar mass relation does not follow the characteristic U-shape with stellar mass. \cite{Davies19b} investigated the fraction of quiescent or passive galaxies in different mass ranges in more detail. They found a low fraction of passive/quenched galaxies at lower masses (5\% at $10^9 M_\odot$ for satellites). \revthree{The general picture from GAMA for satellite galaxies is that they resemble field galaxies and any differences are subtle; the quenched fraction goes down with mass in both field and group satellite galaxies. }

\begin{figure}
    \centering
    \includegraphics[width=0.49\textwidth]{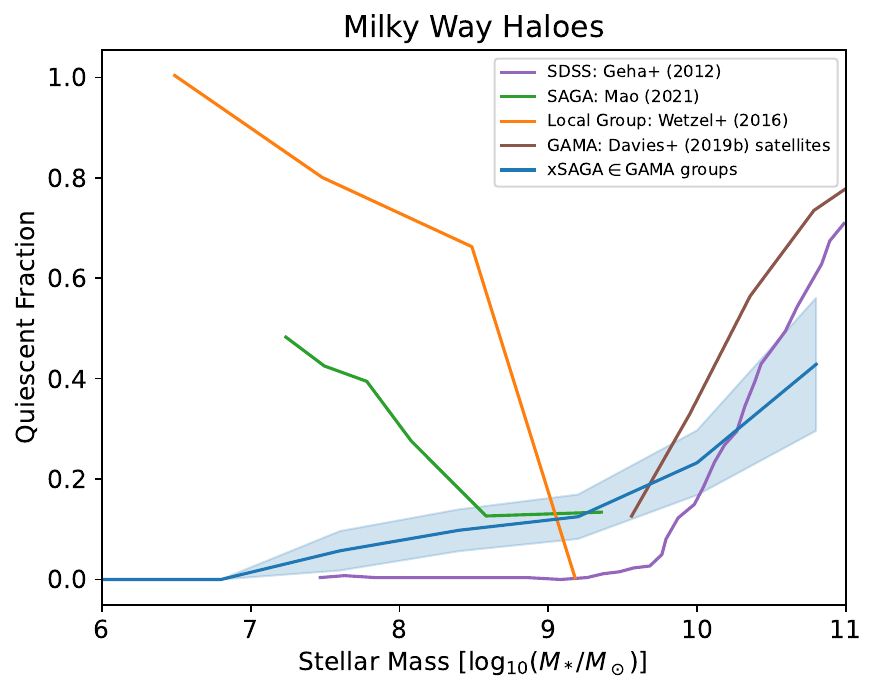}
    \caption{The fraction of quiescent xSAGA galaxies within the R100 of GAMA groups of similar mass as the Milky Way Halo ($\log_\mathrm{10}(M_\mathrm{halo}/M_\odot) = 12-12.4$). An xSAGA galaxy is considered quiescent if the $g-r$ colour is below 0.65 \protect\citep{Salim14a} and their stellar mass is inferred from the $g-r$ colour, the group's redshift, and the apparent sdss-r luminosity following the prescription in \protect\cite{Zibetti09a}. The quiescent fraction is very similar to the one found for GAMA satellites in \protect\cite{Davies19b}}
    \label{f:xSAGA:SFfrac}
\end{figure}

\begin{figure}
    \centering
    \includegraphics[width=0.5\textwidth]{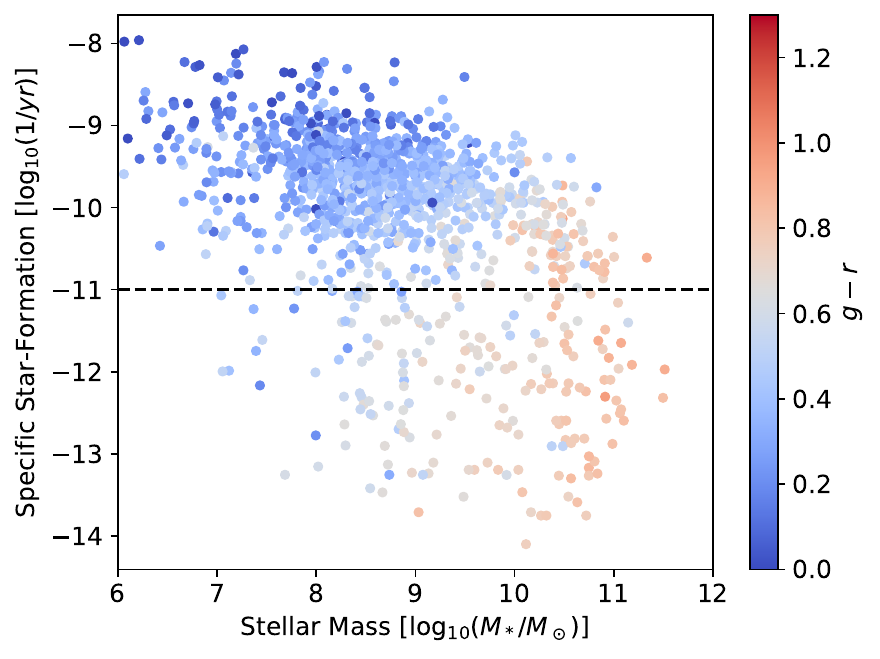}
    \caption{The stellar mass and star-formation rate from MAGPHYS \citep{da-Cunha08} Spectral Energy Distribution fits of the GAMA full photometry \citep{Wright16,Driver22} for the xSAGA galaxies in the GAMA survey. The fraction of star-forming galaxies is determined using the $sSFR < -11 ~ \rm yr^{-1}$ criterion, similar to xSAGA \protect\citep{Geha17}. The colour scale is the $g-r$ colour from the xSAGA catalogue. The colour criterion ($g-r=0.65$) to separate star-forming from quiescent in \protect\cite{Salim14a} is similar but not an exact criterion.  }
    \label{f:xSAGA:GAMA:M-SF}
\end{figure}

In Figure \ref{f:xSAGA:GAMA:M-SF}, we show the stellar mass and specific star-formation \rev{rates, inferred from full GAMA photometry \citep{Wright16,Driver22} using {\sc magphys}} \citep{da-Cunha08}, of the GAMA galaxies that are in the xSAGA catalogue. The dashed line shows the cutoff between star-forming and quiescent used in SAGA. \cite{Salim14a} compared quiescent/star-forming criteria and for the $g-r$ colour, their Figure 4 shows a criterion at $g-r=0.65$. This relation mostly holds for the xSAGA galaxies in GAMA and can be substituted in the xSAGA catalogue as a whole to distinguish between star-forming and quiescent populations. 

Figure \ref{f:xSAGA:SFfrac} shows the fraction of quiescent/passive/quenched galaxies as a function of galaxy stellar mass for the xSAGA galaxies within the R100 radius of known GAMA groups that are similar to the Local Group halo. Figure \ref{f:xSAGA:SFfrac} shows for comparison the quiescent fractions from SDSS \citep{Geha12}, the SAGA survey \citep{Mao21}, the Local Group \citep{Wetzel16}, and all GAMA groups \citep{Davies19b}. 

The passive/quiescent fraction in xSAGA satellites follows the relation by \cite{Davies19b} for GAMA groups for higher satellite galaxy masses ($\log_\mathrm{10} (M_*/M_\odot) > 10$) and the relation found by \cite{Geha12}. \revthree{The lower mass galaxies have increasingly lower quiescent fractions.} 
\revtwo{The useful limit of the early results \citep{Geha12,Davies19b} is $\log(M_*/M_\odot) = 9.5$, below which the surveyed population is not sufficient in size to determine a quiescent fraction. The Local Group mass function is complete to $\log(M_*/M_\odot) = 6$ \citep{Wetzel16} and the targeted SAGA survey \citep{Mao21} to about $\log(M_*/M_\odot) = 7$. The xSAGA search in GAMA groups extends the mass limit down to $\log(M_*/M_\odot) \sim 9$. }

The xSAGA passive fraction bridges to the SAGA passive fractions found at $\log_\mathrm{10} (M_*/M_\odot) \sim 9$. Stellar mass for all xSAGA galaxies is derived using the M/L ratio from the $g-r$ colour \citep{Zibetti09a} and the sdss-r luminosity and the associated group's redshift. Incompleteness increasingly plays a role at the lower mass range and \cite{Wu22d} note their selection completeness and purity corrections are unconstrained below a stellar mass of $\log_\mathrm{10} (M_*/M_\odot) = 7.5$.
We note that xSAGA extends the useful range of GAMA for satellite quiescent fraction about an order of magnitude lower in stellar mass.

\subsection{Local Group Equivalent xSAGA Count}

\begin{figure}
    \centering
    \includegraphics[width=0.5\textwidth]{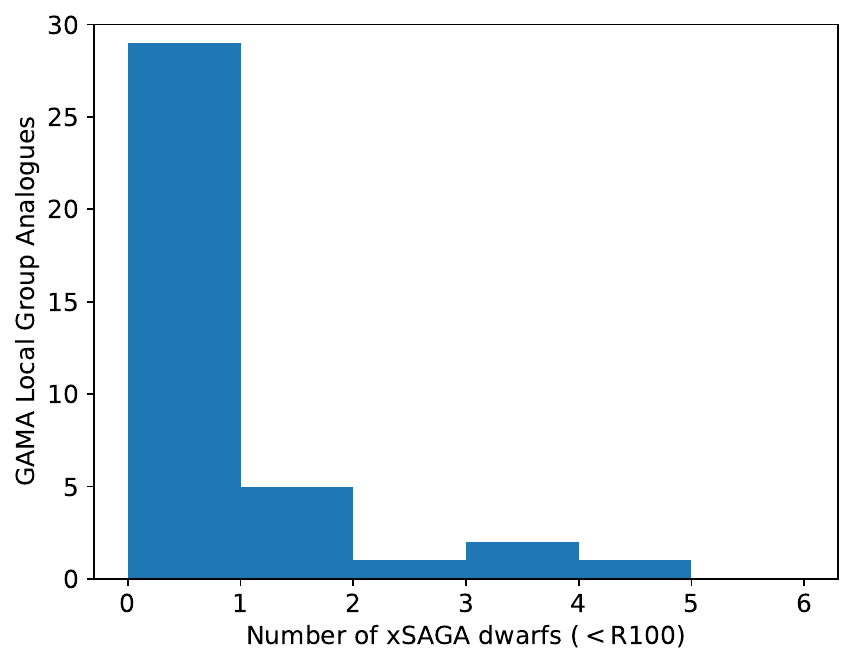}
    \caption{The histogram of the number of xSAGA galaxies, not included in GAMA ($m_r < 21$), within the R100 radius of \rev{the 38} GAMA groups that resemble the Local Group ($12 < \log_\mathrm{10}(M_\mathrm{group}) < 12.5$ and $N_\mathrm{fof} < 4$). Typical number of additional satellites is 1-2.}
    \label{f:LC:hist}
\end{figure}

An extant problem in extragalactic research remains how typical the Local Group of galaxies is compared to other groups of galaxies. 
Figure \ref{f:LC:hist} shows the distribution of the number of xSAGA galaxies, not included in the GAMA galaxy catalogue but within the radius of GAMA groups that resemble the Local Group ($12 < \log_\mathrm{10}(M_\mathrm{group}/M_\odot) < 12.5$ and $N_\mathrm{fof} < 4$), \rev{where $M_\mathrm{group}$ is the dynamical mass and $N_\mathrm{fof}$ the number of members identified in the friends-of-friends algorithm}. There are 38 GAMA groups that fit this criterion. 
The typical number of new small satellites identified by xSAGA appears to be 1-2.

\begin{figure}
    \centering
    \includegraphics[width=0.5\textwidth]{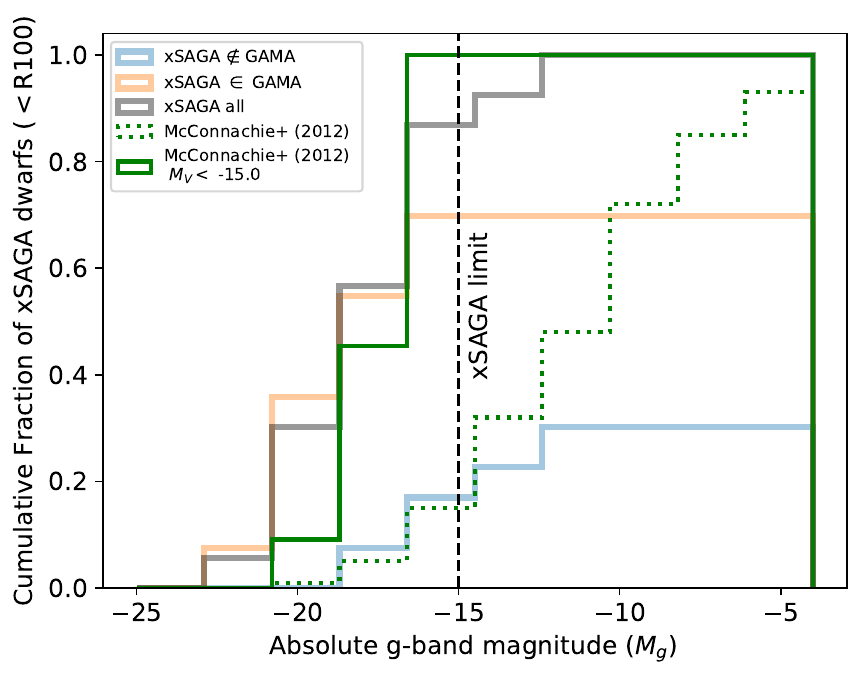}
    \caption{The \rev{cumulative histogram of the} normalized number of xSAGA galaxies in GAMA groups that resemble the Local Group ($12 < \log_\mathrm{10}(M_\mathrm{group}) < 12.5$ and $N_\mathrm{fof} < 4$, \rev{28 unique GAMA groups with 53 xSAGA galaxies in total, 37 xSAGA$\in$GAMA, 16 xSAGA$\notin$GAMA)} as a function of absolute SDSS-g magnitude. For those xSAGA galaxies included in GAMA (orange), we used the GAMA spectroscopic redshift for the distance modulus, for those xSAGA galaxies not in GAMA (blue), we employed the group's spectroscopic redshift ($z_\mathrm{fof}$) for the distance modulus. \rev{The combined xSAGA number is shown as the thick gray line. }
    For comparison, the relative Local Group satellite frequency as a function of V-band luminosity from \protect\cite{McConnachie12} is shown in green. \rev{The observed luminosity function is shown as the dotted line and the cumulative distribution to $M_{lim} =-15$ (dashed black vertical line), the limit of xSAGA is the solid green line.} 
    \rev{The Local Group has relatively fewer bright satellites compared to similar groups in GAMA. Close to the xSAGA limit, the Local Group is richer than the average GAMA group. We note that the completeness of xSAGA may play a role \protect\citep[$\sim$75\% at $M_{lim}=-15$][]{Wu22d}.}}    
    \label{f:LC:LF}
\end{figure}

\begin{table}
    \centering
    \begin{tabular}{l l l l}
Sample            & N xSAGA    & K-S (p-value) & A-D statistic (sig. level) \\
                  & galaxies &   & \\
                  & ($M_{lim} < -15$) &   & \\
\hline
xSAGA$\in$GAMA    &   37 (36)& 0.44 (0.05) & 2.56 (0.03) \\
xSAGA$\notin$GAMA &   16 (7)& 0.27 (0.83) & -0.33 (0.25) \\
xSAGA             &   53 (43)& 0.37 (0.14) & 1.90 (0.05) \\
    \end{tabular}
    \caption{The Kolmogorov-Smirnov and Anderson-Darling statistic of the satellite luminosity function of the Local group and analogue groups with xSAGA satellites, both those xSAGA galaxies already in GAMA, those xSAGA galaxies not in GAMA, and all combined. The number of xSAGA galaxies brighter than the limiting absolute magnitude are listed between brackets.}
    \label{t:LG:stats}
\end{table}

Figure \ref{f:LC:LF} shows the luminosity function of new xSAGA galaxies within GAMA groups (xSAGA $\notin$ GAMA), those xSAGA galaxies that are also in GAMA (xSAGA $\in$ GAMA), and the luminosity function from \cite{McConnachie12} for the Local Group, \rev{both observed and limited to xSAGA observational limit}. 
The xSAGA addition to GAMA groups allows one to examine the brightest satellites of Local Group analogues thanks to the limiting magnitude of xSAGA ($m_r=21AB$ at z=0.03, dashed vertical line). The addition of xSAGA allows for satellites a magnitude fainter than GAMA to be included in group statistics.  
CNN projects like xSAGA will have to be pushed to lower luminosities with future imaging surveys to fully probe the Local Group luminosity function. 

\rev{Figure \ref{f:LC:LF} compares the Local Group relative numbers to those of similar dynamical mass and membership in GAMA \rev{using a cumulative, normalized histogram.} \revthree{We show the full range of Local Group luminosity function to illustrate the length current deep imaging and spectroscopic still have to go before fully sampling a similar range in luminosities.} Normalized over the total number of satellites found, the GAMA groups have a \rev{relatively} more bright satellites than the Local Group, the majority of which were already known in the GAMA catalogue \revthree{($M_g < -17$)}. Adding xSAGA adds number at the intermediate absolute magnitudes ($-17< M_g < -13$). \revthree{Because the xSAGA selection did not include a hard limiting magnitude, the number of xSAGA satellites added still rises beyong $M_g=-15$, hinting at the dimmer populations yet to be discovered and confirmed. }
The Local Group is very similar to these GAMA groups, when limited to the same absolute magnitude. The Kolmogorov-Smirnov and Anderson-Darling statistical measures whether both luminosity functions above $M_{lim}=-15$ are from the same distribution are inconclusive, i.e. the null hypothesis of originating from the same distribution cannot be rejected with any confidence see Table \ref{t:LG:stats}).}

\revtwo{Previously, \cite{Guo11c} and \cite{Jiang12a} reported a factor two fewer extra-galactic satellites than the average of MW and M31 satellites to a limit of $M_V < -15$, based on the SDSS and CFHT imaging survey respectively. However, \cite{Wang21i} concluded that the number of satellites match for Local Group analogues based on the HyperSuprimeCam, DECaLS, and SDSS imaging surveys. A logical following step is to expand the search for the lowest mass galaxies in nearby groups of galaxies \citep{Tanaka18,Mutlu-Pakdil22}. For example, the M81 grouping \citep{Chiboucas13,Monachesi14} or the M101 \citep{Danieli17,Bennet20,Garner22} as nearby analogues for the Milky Way and its satellite retinue. Deeper imaging surveys have long held the promise to expand the satellite mass or luminosity functions of nearby galaxies \citep{Tollerud08,Danieli18,Wang21i}, the Rubin Observatory certainly promises to do so \citep{Mutlu-Pakdil21}. Once greater statistics are obtained on the satellite retinue of Milky Way/Local Group analogues \citep{Busha11,DSouza14}, the Milky Way and its satellites can be placed in an evolutionary context \citep{Lan16,Danieli22}. Such group environment statistics can then be compared to, for example, completely isolated populations of low-mass galaxies \citep{de-los-Reyes23a}, to discern the effects of environment on low mass galaxy evolution.}

\begin{figure}
    \centering
    \includegraphics[width=0.49\textwidth]{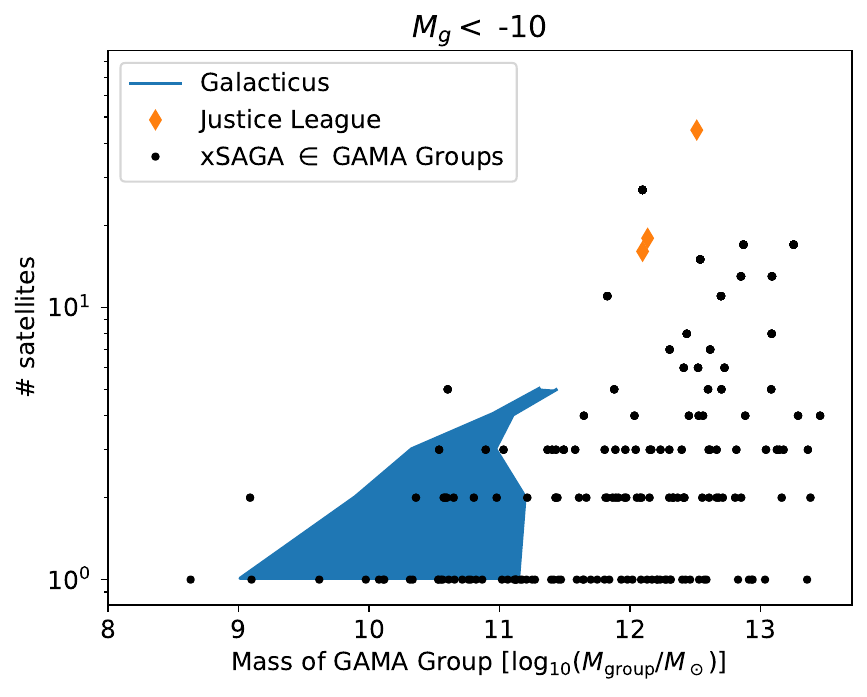}
    \caption{The number of satellites in a virialized halo for the Galacticus \protect\citep[blue filled area][Bovill \textit{private communication}]{Benson12,Brennan19} and Justice League \protect\citep[orange diamond][]{Applebaum21,Akins21} simulations of group galaxies for satellites of absolute magnitudes of $M_V < -10$ mag, the practical limit of xSAGA.  }
    \label{f:Mvir:Nsat}
\end{figure}

Figure \ref{f:Mvir:Nsat} shows the numbers of satellites in two suites of simulations, the \rev{semi-analytical} Galacticus \citep[][]{Brennan19, Weerasooriya23} and \rev{smoothed particle hydrodynamical models of} the Justice League \citep{Applebaum21,Akins21} for different luminosity cuts of the satellite population. \rev{The plot compares the results of the computationally expensive Justice League results to the much more computationally inexpensive Galacticus. The Justice League runs benefit from detailed physics at the highest resolutions presently run while the Galacticus simulations can be thought of as a Local Group grounding of the prescriptions that make up semi-analytical models (Bovill et al. {\it in prep.}).} 

Figure \ref{f:Mvir:Nsat}  shows the number of xSAGA galaxies --both those already in GAMA and new-- \revthree{with an absolute luminosity of $M_g < -10$,} within the R100 radius for unique GAMA groups in our sample. The numbers are consistent \rev{for the intermediate mass groups ($10 < \log_{10}(M_\mathrm{group}/M_\odot) < 11$) with Galacticus but not for lower mass groups ($\log_{10}(M_\mathrm{group}/M_\odot) < 10$)
for Galacticus. For the higher mass groups ($\log_{10}(M_\mathrm{group}/M_\odot) \sim 12$) the top satellite numbers agree well with the Justice League ones. }

\section{Conclusions}
\label{s:conclusions}

In this paper, we report a simple experiment where we compare the positions of very likely low redshift ($z<0.03$, $P_\mathrm{z<0.03} >$ 50\%) galaxies identified with a CNN on DESI data \citep{Wu22d} to the positions of galaxy groups identified in the Galaxy And Mass Assembly survey using a FoF algorithm \citep{Robotham11,Robotham12}. Exploring the number of likely low-redshift galaxies associated with GAMA groups, we find that the number of satellites associated with a GAMA group is between none and five for the majority of local GAMA groups. As can be expected, more massive and richer groups show a larger number of associated xSAGA satellites. Most of the associated xSAGA satellites span a range of surface brightness, are of intermediate colour ($0<g-r<1$) and small ($R_\mathrm{eff} < 10$ kpc). While the xSAGA galaxies add a number of new members to groups, their total stellar mass contribution to the group is very small: typically less than 1\% for the \rev{new} xSAGA galaxies in a GAMA group.

We checked the redshifts of the xSAGA selection against the spectroscopic redshifts from GAMA and found a false positive rate of 30\% for the selection, consistent with \cite{Wu22d} predictions (Figure \ref{f:xsaga:gama}). The xSAGA extension of the low-redshift GAMA groups constitutes a probabilistic extension to the group catalogue membership. This looks to be a promising approach to supplement high-spectroscopic completeness surveys with probabilistic galaxy catalogues to attain an ever more complete picture of groups of galaxies, not just the retinue of Milky Way equivalent galaxies.

We compare the numbers of satellites found with xSAGA to group mass (Figure \ref{f:Mvir:Nsat} and these are in qualitative agreement with the Galacticus \citep{Brennan19} and Justice League \citep{Applebaum21,Akins21} simulations of galaxy groups. 

Selecting those GAMA groups that resemble the Local Group of galaxies ($12 < \log_\mathrm{10}(M_\mathrm{group}/M_\odot) < 12.5$ and $N_\mathrm{fof} < 4$), we find that such groups have a typical retinue of 1-2 xSAGA galaxies. \revtwo{The satellite luminosity function of Local Group analogue GAMA groups is quite similar to the one observed in the Local Group down to the equivalent limit of absolute luminosity for xSAGA (Figure \ref{f:LC:LF}).} Starting from those xSAGA galaxies that were included in GAMA observations, we explore the quiescent fraction of xSAGA galaxies as a function of stellar mass. This fraction is consistent with the one found in previous work \citep{Geha12,Davies19b} but extends the useable range of stellar masses down to the SAGA survey's \revtwo{(Figure \ref{f:xSAGA:GAMA:M-SF})}. 

In this paper, we show that the xSAGA extension using a CNN by \cite{Wu22d} is a very useful tool to expand a catalogue of known galaxy groups to lower stellar galaxy masses and we confirm independently the reported purity of the xSAGA sample (70\%) with GAMA spectroscopic redshifts. The xSAGA expansion can be used for spectroscopic target pre-selection and statistical comparison between group memberships and simulations of galaxy groups.

\section*{Data Availability}

The GAMA DR4 \citep{Driver22} included v10 of the group catalogue {\sc gamagroupv10} which describes the GAMA II equatorial and G02 survey regions. 
The full xSAGA satellite catalogue is not yet available but is expected to appear shortly.

\section*{Acknowledgements}

BWH would like to thank Karl Gordon for hosting him at STSCI during which visit, this idea crystallized. Special thanks to John Wu for making the xSAGA catalogue available and answering questions on his work. 





\begin{thebibliography}{}
\makeatletter
\relax
\def\mn@urlcharsother{\let\do\@makeother \do\$\do\&\do\#\do\^\do\_\do\%\do\~}
\def\mn@doi{\begingroup\mn@urlcharsother \@ifnextchar [ {\mn@doi@}
  {\mn@doi@[]}}
\def\mn@doi@[#1]#2{\def\@tempa{#1}\ifx\@tempa\@empty \href
  {http://dx.doi.org/#2} {doi:#2}\else \href {http://dx.doi.org/#2} {#1}\fi
  \endgroup}
\def\mn@eprint#1#2{\mn@eprint@#1:#2::\@nil}
\def\mn@eprint@arXiv#1{\href {http://arxiv.org/abs/#1} {{\tt arXiv:#1}}}
\def\mn@eprint@dblp#1{\href {http://dblp.uni-trier.de/rec/bibtex/#1.xml}
  {dblp:#1}}
\def\mn@eprint@#1:#2:#3:#4\@nil{\def\@tempa {#1}\def\@tempb {#2}\def\@tempc
  {#3}\ifx \@tempc \@empty \let \@tempc \@tempb \let \@tempb \@tempa \fi \ifx
  \@tempb \@empty \def\@tempb {arXiv}\fi \@ifundefined
  {mn@eprint@\@tempb}{\@tempb:\@tempc}{\expandafter \expandafter \csname
  mn@eprint@\@tempb\endcsname \expandafter{\@tempc}}}

\bibitem[\protect\citeauthoryear{{Akins}, {Christensen}, {Brooks}, {Munshi},
  {Applebaum}, {Engelhardt}  \& {Chamberland}}{{Akins} et~al.}{2021}]{Akins21}
{Akins} H.~B.,  {Christensen} C.~R.,  {Brooks} A.~M.,  {Munshi} F.,
  {Applebaum} E.,  {Engelhardt} A.,   {Chamberland} L.,  2021, \apj, 909, 139

\bibitem[\protect\citeauthoryear{{Alpaslan} et~al.,}{{Alpaslan}
  et~al.}{2014}]{Alpaslan14}
{Alpaslan} M.,  et~al., 2014, \mnras, 440, L106

\bibitem[\protect\citeauthoryear{{Alpaslan} et~al.,}{{Alpaslan}
  et~al.}{2015}]{Alpaslan15}
{Alpaslan} M.,  et~al., 2015, \mnras, 451, 3249

\bibitem[\protect\citeauthoryear{{Applebaum}, {Brooks}, {Christensen},
  {Munshi}, {Quinn}, {Shen}  \& {Tremmel}}{{Applebaum}
  et~al.}{2021}]{Applebaum21}
{Applebaum} E.,  {Brooks} A.~M.,  {Christensen} C.~R.,  {Munshi} F.,  {Quinn}
  T.~R.,  {Shen} S.,   {Tremmel} M.,  2021, \apj, 906, 96

\bibitem[\protect\citeauthoryear{{Baldry} et~al.,}{{Baldry}
  et~al.}{2018}]{Baldry18}
{Baldry} I.~K.,  et~al., 2018, \mnras, 474, 3875

\bibitem[\protect\citeauthoryear{{Bechtol} et~al.,}{{Bechtol}
  et~al.}{2015}]{Bechtol15}
{Bechtol} K.,  et~al., 2015, \apj, 807, 50

\bibitem[\protect\citeauthoryear{{Bennet}, {Sand}, {Crnojevi{\'c}}, {Spekkens},
  {Karunakaran}, {Zaritsky}  \& {Mutlu-Pakdil}}{{Bennet}
  et~al.}{2020}]{Bennet20}
{Bennet} P.,  {Sand} D.~J.,  {Crnojevi{\'c}} D.,  {Spekkens} K.,  {Karunakaran}
  A.,  {Zaritsky} D.,   {Mutlu-Pakdil} B.,  2020, \apjl, 893, L9

\bibitem[\protect\citeauthoryear{{Benson}}{{Benson}}{2012}]{Benson12}
{Benson} A.~J.,  2012, \na, 17, 175

\bibitem[\protect\citeauthoryear{{Benson}, {Frenk}, {Lacey}, {Baugh}  \&
  {Cole}}{{Benson} et~al.}{2002}]{Benson02}
{Benson} A.~J.,  {Frenk} C.~S.,  {Lacey} C.~G.,  {Baugh} C.~M.,   {Cole} S.,
  2002, \mnras, 333, 177

\bibitem[\protect\citeauthoryear{{Boylan-Kolchin}, {Besla}  \&
  {Hernquist}}{{Boylan-Kolchin} et~al.}{2011}]{Boylan-Kolchin11}
{Boylan-Kolchin} M.,  {Besla} G.,   {Hernquist} L.,  2011, \mnras, 414, 1560

\bibitem[\protect\citeauthoryear{{Boylan-Kolchin}, {Weisz}, {Bullock}  \&
  {Cooper}}{{Boylan-Kolchin} et~al.}{2016}]{Boylan-Kolchin16}
{Boylan-Kolchin} M.,  {Weisz} D.~R.,  {Bullock} J.~S.,   {Cooper} M.~C.,  2016,
  \mnras, 462, L51

\bibitem[\protect\citeauthoryear{{Brennan}, {Benson}, {Cyr-Racine}, {Keeton},
  {Moustakas}  \& {Pullen}}{{Brennan} et~al.}{2019}]{Brennan19}
{Brennan} S.,  {Benson} A.~J.,  {Cyr-Racine} F.-Y.,  {Keeton} C.~R.,
  {Moustakas} L.~A.,   {Pullen} A.~R.,  2019, \mnras, 488, 5085

\bibitem[\protect\citeauthoryear{{Brough} et~al.,}{{Brough}
  et~al.}{2011}]{Brough11a}
{Brough} S.,  et~al., 2011, \mnras, 413, 1236

\bibitem[\protect\citeauthoryear{{Busha}, {Wechsler}, {Behroozi}, {Gerke},
  {Klypin}  \& {Primack}}{{Busha} et~al.}{2011}]{Busha11}
{Busha} M.~T.,  {Wechsler} R.~H.,  {Behroozi} P.~S.,  {Gerke} B.~F.,  {Klypin}
  A.~A.,   {Primack} J.~R.,  2011, \apj, 743, 117

\bibitem[\protect\citeauthoryear{{Carlsten}, {Greene}, {Greco}, {Beaton}  \&
  {Kado-Fong}}{{Carlsten} et~al.}{2021}]{Carlsten21}
{Carlsten} S.~G.,  {Greene} J.~E.,  {Greco} J.~P.,  {Beaton} R.~L.,
  {Kado-Fong} E.,  2021, \apj, 922, 267

\bibitem[\protect\citeauthoryear{{Carlsten}, {Greene}, {Beaton}  \&
  {Greco}}{{Carlsten} et~al.}{2022a}]{Carlsten22b}
{Carlsten} S.~G.,  {Greene} J.~E.,  {Beaton} R.~L.,   {Greco} J.~P.,  2022a,
  \apj, 927, 44

\bibitem[\protect\citeauthoryear{{Carlsten}, {Greene}, {Beaton}, {Danieli}  \&
  {Greco}}{{Carlsten} et~al.}{2022b}]{Carlsten22a}
{Carlsten} S.~G.,  {Greene} J.~E.,  {Beaton} R.~L.,  {Danieli} S.,   {Greco}
  J.~P.,  2022b, \apj, 933, 47

\bibitem[\protect\citeauthoryear{{Casey}, {Greco}, {Peter}  \& {Davis}}{{Casey}
  et~al.}{2023}]{Casey23}
{Casey} K.~J.,  {Greco} J.~P.,  {Peter} A. H.~G.,   {Davis} A.~B.,  2023,
  \mnras, 520, 4715

\bibitem[\protect\citeauthoryear{{Chiboucas}, {Jacobs}, {Tully}  \&
  {Karachentsev}}{{Chiboucas} et~al.}{2013}]{Chiboucas13}
{Chiboucas} K.,  {Jacobs} B.~A.,  {Tully} R.~B.,   {Karachentsev} I.~D.,  2013,
  \aj, 146, 126

\bibitem[\protect\citeauthoryear{{Collins}, {Rich}  \& {Chapman}}{{Collins}
  et~al.}{2012}]{Collins12}
{Collins} M.~L.~M.,  {Rich} R.~M.,   {Chapman} S.~C.,  2012, in {Aoki} W.,
  {Ishigaki} M.,  {Suda} T.,  {Tsujimoto} T.,   {Arimoto} N.,  eds,
  Astronomical Society of the Pacific Conference Series Vol. 458, Galactic
  Archaeology: Near-Field Cosmology and the Formation of the Milky Way. p.~319

\bibitem[\protect\citeauthoryear{{Collins} et~al.,}{{Collins}
  et~al.}{2014}]{Collins14a}
{Collins} M. L.~M.,  et~al., 2014, \apj, 783, 7

\bibitem[\protect\citeauthoryear{{Collins}, {Rich}, {Ibata}, {Martin}  \&
  {Preston}}{{Collins} et~al.}{2016}]{Collins16}
{Collins} M.~L.~M.,  {Rich} R.~M.,  {Ibata} R.~A.,  {Martin} N.~F.,   {Preston}
  J.,  2016, preprint

\bibitem[\protect\citeauthoryear{{Creasey}, {Scannapieco}, {Nuza}, {Yepes},
  {Gottl{\"o}ber}  \& {Steinmetz}}{{Creasey} et~al.}{2015}]{Creasey15}
{Creasey} P.,  {Scannapieco} C.,  {Nuza} S.~E.,  {Yepes} G.,  {Gottl{\"o}ber}
  S.,   {Steinmetz} M.,  2015, \apjl, 800, L4

\bibitem[\protect\citeauthoryear{{D'Souza}, {Kauffman}, {Wang}  \&
  {Vegetti}}{{D'Souza} et~al.}{2014}]{DSouza14}
{D'Souza} R.,  {Kauffman} G.,  {Wang} J.,   {Vegetti} S.,  2014, \mnras, 443,
  1433

\bibitem[\protect\citeauthoryear{{Danieli}, {van Dokkum}, {Merritt}, {Abraham},
  {Zhang}, {Karachentsev}  \& {Makarova}}{{Danieli} et~al.}{2017}]{Danieli17}
{Danieli} S.,  {van Dokkum} P.,  {Merritt} A.,  {Abraham} R.,  {Zhang} J.,
  {Karachentsev} I.~D.,   {Makarova} L.~N.,  2017, \apj, 837, 136

\bibitem[\protect\citeauthoryear{{Danieli}, {van Dokkum}  \&
  {Conroy}}{{Danieli} et~al.}{2018}]{Danieli18}
{Danieli} S.,  {van Dokkum} P.,   {Conroy} C.,  2018, \apj, 856, 69

\bibitem[\protect\citeauthoryear{{Danieli}, {Greene}, {Carlsten}, {Jiang},
  {Beaton}  \& {Goulding}}{{Danieli} et~al.}{2022}]{Danieli22}
{Danieli} S.,  {Greene} J.~E.,  {Carlsten} S.,  {Jiang} F.,  {Beaton} R.,
  {Goulding} A.~D.,  2022, arXiv e-prints, p. arXiv:2210.14233

\bibitem[\protect\citeauthoryear{{Davies} et~al.,}{{Davies}
  et~al.}{2019a}]{Davies19a}
{Davies} L.~J.~M.,  et~al., 2019a, \mnras, 483, 1881

\bibitem[\protect\citeauthoryear{{Davies} et~al.,}{{Davies}
  et~al.}{2019b}]{Davies19b}
{Davies} L.~J.~M.,  et~al., 2019b, \mnras, 483, 5444

\bibitem[\protect\citeauthoryear{{Dey} et~al.,}{{Dey} et~al.}{2019}]{Dey19}
{Dey} A.,  et~al., 2019, \aj, 157, 168

\bibitem[\protect\citeauthoryear{{Digby} et~al.,}{{Digby}
  et~al.}{2019}]{Digby19}
{Digby} R.,  et~al., 2019, \mnras, 485, 5423

\bibitem[\protect\citeauthoryear{{Driver} et~al.,}{{Driver}
  et~al.}{2009}]{Driver09}
{Driver} S.~P.,  et~al., 2009, Astronomy and Geophysics, 50, 050000

\bibitem[\protect\citeauthoryear{{Driver} et~al.,}{{Driver}
  et~al.}{2011}]{Driver11}
{Driver} S.~P.,  et~al., 2011, \mnras, 413, 971

\bibitem[\protect\citeauthoryear{{Driver} et~al.,}{{Driver}
  et~al.}{2016}]{Driver16}
{Driver} S.~P.,  et~al., 2016, \apj, 827, 108

\bibitem[\protect\citeauthoryear{{Driver} et~al.,}{{Driver}
  et~al.}{2022}]{Driver22}
{Driver} S.~P.,  et~al., 2022, \mnras, 513, 439

\bibitem[\protect\citeauthoryear{{Drlica-Wagner} et~al.,}{{Drlica-Wagner}
  et~al.}{2015}]{Drlica-Wagner15}
{Drlica-Wagner} A.,  et~al., 2015, \apj, 813, 109

\bibitem[\protect\citeauthoryear{{Drlica-Wagner} et~al.,}{{Drlica-Wagner}
  et~al.}{2020}]{Drlica-Wagner20}
{Drlica-Wagner} A.,  et~al., 2020, \apj, 893, 47

\bibitem[\protect\citeauthoryear{{Eke} et~al.,}{{Eke} et~al.}{2004}]{Eke04}
{Eke} V.~R.,  et~al., 2004, \mnras, 355, 769

\bibitem[\protect\citeauthoryear{{Elias}, {Sales}, {Creasey}, {Cooper},
  {Bullock}, {Rich}  \& {Hernquist}}{{Elias} et~al.}{2018}]{Elias18}
{Elias} L.~M.,  {Sales} L.~V.,  {Creasey} P.,  {Cooper} M.~C.,  {Bullock}
  J.~S.,  {Rich} R.~M.,   {Hernquist} L.,  2018, \mnras, 479, 4004

\bibitem[\protect\citeauthoryear{{Fattahi}, {Navarro}, {Starkenburg}, {Barber}
  \& {McConnachie}}{{Fattahi} et~al.}{2013}]{Fattahi13}
{Fattahi} A.,  {Navarro} J.~F.,  {Starkenburg} E.,  {Barber} C.~R.,
  {McConnachie} A.~W.,  2013, \mnras, 431, L73

\bibitem[\protect\citeauthoryear{{Fattahi} et~al.,}{{Fattahi}
  et~al.}{2016}]{Fattahi16}
{Fattahi} A.,  et~al., 2016, \mnras, 457, 844

\bibitem[\protect\citeauthoryear{{Forero-Romero}, {Hoffman}, {Yepes},
  {Gottl{\"o}ber}, {Piontek}, {Klypin}  \& {Steinmetz}}{{Forero-Romero}
  et~al.}{2011}]{Forero-Romero11}
{Forero-Romero} J.~E.,  {Hoffman} Y.,  {Yepes} G.,  {Gottl{\"o}ber} S.,
  {Piontek} R.,  {Klypin} A.,   {Steinmetz} M.,  2011, \mnras, 417, 1434

\bibitem[\protect\citeauthoryear{{Garling}, {Peter}, {Kochanek}, {Sand}  \&
  {Crnojevi{\'c}}}{{Garling} et~al.}{2020}]{Garling20}
{Garling} C.~T.,  {Peter} A. H.~G.,  {Kochanek} C.~S.,  {Sand} D.~J.,
  {Crnojevi{\'c}} D.,  2020, \mnras, 492, 1713

\bibitem[\protect\citeauthoryear{{Garner}, {Mihos}, {Harding}, {Watkins}  \&
  {McGaugh}}{{Garner} et~al.}{2022}]{Garner22}
{Garner} R.,  {Mihos} J.~C.,  {Harding} P.,  {Watkins} A.~E.,   {McGaugh}
  S.~S.,  2022, \apj, 941, 182

\bibitem[\protect\citeauthoryear{{Garrison-Kimmel}, {Boylan-Kolchin}, {Bullock}
   \& {Kirby}}{{Garrison-Kimmel} et~al.}{2014}]{Garrison-Kimmel14}
{Garrison-Kimmel} S.,  {Boylan-Kolchin} M.,  {Bullock} J.~S.,   {Kirby} E.~N.,
  2014, \mnras, 444, 222

\bibitem[\protect\citeauthoryear{{Garrison-Kimmel} et~al.,}{{Garrison-Kimmel}
  et~al.}{2019a}]{Garrison-Kimmel19a}
{Garrison-Kimmel} S.,  et~al., 2019a, \mnras, 487, 1380

\bibitem[\protect\citeauthoryear{{Garrison-Kimmel} et~al.,}{{Garrison-Kimmel}
  et~al.}{2019b}]{Garrison-Kimmel19b}
{Garrison-Kimmel} S.,  et~al., 2019b, \mnras, 489, 4574

\bibitem[\protect\citeauthoryear{{Geha}, {Blanton}, {Yan}  \& {Tinker}}{{Geha}
  et~al.}{2012}]{Geha12}
{Geha} M.,  {Blanton} M.~R.,  {Yan} R.,   {Tinker} J.~L.,  2012, \apj, 757, 85

\bibitem[\protect\citeauthoryear{{Geha} et~al.,}{{Geha} et~al.}{2017}]{Geha17}
{Geha} M.,  et~al., 2017, preprint

\bibitem[\protect\citeauthoryear{{Gottloeber}, {Hoffman}  \&
  {Yepes}}{{Gottloeber} et~al.}{2010}]{Gottloeber10}
{Gottloeber} S.,  {Hoffman} Y.,   {Yepes} G.,  2010, arXiv e-prints, p.
  arXiv:1005.2687

\bibitem[\protect\citeauthoryear{{Grootes} et~al.,}{{Grootes}
  et~al.}{2017}]{Grootes17}
{Grootes} M.~W.,  et~al., 2017, \aj, 153, 111

\bibitem[\protect\citeauthoryear{{Grootes} et~al.,}{{Grootes}
  et~al.}{2018}]{Grootes18}
{Grootes} M.~W.,  et~al., 2018, \mnras, 477, 1015

\bibitem[\protect\citeauthoryear{{Guo}, {Cole}, {Eke}  \& {Frenk}}{{Guo}
  et~al.}{2011}]{Guo11c}
{Guo} Q.,  {Cole} S.,  {Eke} V.,   {Frenk} C.,  2011, \mnras, 417, 370

\bibitem[\protect\citeauthoryear{{Haines}, {Gargiulo}, {La Barbera},
  {Mercurio}, {Merluzzi}  \& {Busarello}}{{Haines} et~al.}{2007}]{Haines07}
{Haines} C.~P.,  {Gargiulo} A.,  {La Barbera} F.,  {Mercurio} A.,  {Merluzzi}
  P.,   {Busarello} G.,  2007, \mnras, 381, 7

\bibitem[\protect\citeauthoryear{{Hammer}, {Yang}, {Fouquet}, {Pawlowski},
  {Kroupa}, {Puech}, {Flores}  \& {Wang}}{{Hammer} et~al.}{2013}]{Hammer13}
{Hammer} F.,  {Yang} Y.,  {Fouquet} S.,  {Pawlowski} M.~S.,  {Kroupa} P.,
  {Puech} M.,  {Flores} H.,   {Wang} J.,  2013, \mnras, 431, 3543

\bibitem[\protect\citeauthoryear{{Homma} et~al.,}{{Homma}
  et~al.}{2019}]{Homma19}
{Homma} D.,  et~al., 2019, \pasj, 71, 94

\bibitem[\protect\citeauthoryear{{Jiang}, {Jing}  \& {Li}}{{Jiang}
  et~al.}{2012}]{Jiang12a}
{Jiang} C.~Y.,  {Jing} Y.~P.,   {Li} C.,  2012, \apj, 760, 16

\bibitem[\protect\citeauthoryear{{Kauffmann}, {Li}, {Zhang}  \&
  {Weinmann}}{{Kauffmann} et~al.}{2013}]{Kauffmann13a}
{Kauffmann} G.,  {Li} C.,  {Zhang} W.,   {Weinmann} S.,  2013, \mnras, 430,
  1447

\bibitem[\protect\citeauthoryear{{Kawinwanichakij} et~al.,}{{Kawinwanichakij}
  et~al.}{2017}]{Kawinwanichakij17}
{Kawinwanichakij} L.,  et~al., 2017, \apj, 847, 134

\bibitem[\protect\citeauthoryear{{Kim}, {Jerjen}, {Mackey}, {Da Costa}  \&
  {Milone}}{{Kim} et~al.}{2015}]{Kim15c}
{Kim} D.,  {Jerjen} H.,  {Mackey} D.,  {Da Costa} G.~S.,   {Milone} A.~P.,
  2015, \apjl, 804, L44

\bibitem[\protect\citeauthoryear{{Klypin}, {Zhao}  \& {Somerville}}{{Klypin}
  et~al.}{2002}]{Klypin02}
{Klypin} A.,  {Zhao} H.,   {Somerville} R.~S.,  2002, \apj, 573, 597

\bibitem[\protect\citeauthoryear{{Koposov}, {Yoo}, {Rix}, {Weinberg},
  {Macci{\`o}}  \& {Escud{\'e}}}{{Koposov} et~al.}{2009}]{Koposov09}
{Koposov} S.~E.,  {Yoo} J.,  {Rix} H.-W.,  {Weinberg} D.~H.,  {Macci{\`o}}
  A.~V.,   {Escud{\'e}} J.~M.,  2009, \apj, 696, 2179

\bibitem[\protect\citeauthoryear{{Koposov}, {Belokurov}, {Torrealba}  \&
  {Evans}}{{Koposov} et~al.}{2015}]{Koposov15}
{Koposov} S.~E.,  {Belokurov} V.,  {Torrealba} G.,   {Evans} N.~W.,  2015,
  \apj, 805, 130

\bibitem[\protect\citeauthoryear{{Kravtsov}}{{Kravtsov}}{2002}]{Kravtsov02}
{Kravtsov} V.,  2002, \aap, 396, 117

\bibitem[\protect\citeauthoryear{{Lan}, {M{\'e}nard}  \& {Mo}}{{Lan}
  et~al.}{2016}]{Lan16}
{Lan} T.-W.,  {M{\'e}nard} B.,   {Mo} H.,  2016, \mnras, 459, 3998

\bibitem[\protect\citeauthoryear{{Libeskind}, {Knebe}, {Hoffman},
  {Gottl{\"o}ber}, {Yepes}  \& {Steinmetz}}{{Libeskind}
  et~al.}{2011}]{Libeskind11}
{Libeskind} N.~I.,  {Knebe} A.,  {Hoffman} Y.,  {Gottl{\"o}ber} S.,  {Yepes}
  G.,   {Steinmetz} M.,  2011, \mnras, 411, 1525

\bibitem[\protect\citeauthoryear{{Libeskind} et~al.,}{{Libeskind}
  et~al.}{2020}]{Libeskind20}
{Libeskind} N.~I.,  et~al., 2020, \mnras, 498, 2968

\bibitem[\protect\citeauthoryear{{Liske} et~al.,}{{Liske}
  et~al.}{2015}]{Liske15}
{Liske} J.,  et~al., 2015, \mnras, 452, 2087

\bibitem[\protect\citeauthoryear{{Lovell}, {Eke}, {Frenk}  \&
  {Jenkins}}{{Lovell} et~al.}{2011}]{Lovell11}
{Lovell} M.~R.,  {Eke} V.~R.,  {Frenk} C.~S.,   {Jenkins} A.,  2011, \mnras,
  413, 3013

\bibitem[\protect\citeauthoryear{{Mao}, {Geha}, {Wechsler}, {Weiner},
  {Tollerud}, {Nadler}  \& {Kallivayalil}}{{Mao} et~al.}{2021}]{Mao21}
{Mao} Y.-Y.,  {Geha} M.,  {Wechsler} R.~H.,  {Weiner} B.,  {Tollerud} E.~J.,
  {Nadler} E.~O.,   {Kallivayalil} N.,  2021, \apj, 907, 85

\bibitem[\protect\citeauthoryear{{Mart{\'\i}nez-Delgado}
  et~al.,}{{Mart{\'\i}nez-Delgado} et~al.}{2016}]{Martinez-Delgado16}
{Mart{\'\i}nez-Delgado} D.,  et~al., 2016, \aj, 151, 96

\bibitem[\protect\citeauthoryear{{Mart{\'\i}nez-Lombilla}
  et~al.,}{{Mart{\'\i}nez-Lombilla} et~al.}{2023}]{Martinez-Lombilla23}
{Mart{\'\i}nez-Lombilla} C.,  et~al., 2023, \mnras, 518, 1195

\bibitem[\protect\citeauthoryear{{McConnachie}}{{McConnachie}}{2012}]{McConnachie12}
{McConnachie} A.~W.,  2012, \aj, 144, 4

\bibitem[\protect\citeauthoryear{{Monachesi} et~al.,}{{Monachesi}
  et~al.}{2014}]{Monachesi14}
{Monachesi} A.,  et~al., 2014, \apj, 780, 179

\bibitem[\protect\citeauthoryear{{Mutlu-Pakdil} et~al.,}{{Mutlu-Pakdil}
  et~al.}{2021}]{Mutlu-Pakdil21}
{Mutlu-Pakdil} B.,  et~al., 2021, \apj, 918, 88

\bibitem[\protect\citeauthoryear{{Mutlu-Pakdil} et~al.,}{{Mutlu-Pakdil}
  et~al.}{2022}]{Mutlu-Pakdil22}
{Mutlu-Pakdil} B.,  et~al., 2022, \apj, 926, 77

\bibitem[\protect\citeauthoryear{{Okamoto}, {Frenk}, {Jenkins}  \&
  {Theuns}}{{Okamoto} et~al.}{2010}]{Okamoto10}
{Okamoto} T.,  {Frenk} C.~S.,  {Jenkins} A.,   {Theuns} T.,  2010, \mnras, 406,
  208

\bibitem[\protect\citeauthoryear{{Pawlowski}, {Kroupa}  \&
  {Jerjen}}{{Pawlowski} et~al.}{2013}]{Pawlowski13}
{Pawlowski} M.~S.,  {Kroupa} P.,   {Jerjen} H.,  2013, \mnras, 435, 1928

\bibitem[\protect\citeauthoryear{{Pearson}, {Wang}, {Brough}, {Holwerda},
  {Hopkins}  \& {Loveday}}{{Pearson} et~al.}{2021}]{Pearson21}
{Pearson} W.~J.,  {Wang} L.,  {Brough} S.,  {Holwerda} B.~W.,  {Hopkins} A.~M.,
    {Loveday} J.,  2021, \aap, 646, A151

\bibitem[\protect\citeauthoryear{{Planck Collaboration} et~al.,}{{Planck
  Collaboration} et~al.}{2018}]{Planck-Collaboration18b}
{Planck Collaboration} et~al., 2018, preprint

\bibitem[\protect\citeauthoryear{{Polzin}, {van Dokkum}, {Danieli}, {Greco}  \&
  {Romanowsky}}{{Polzin} et~al.}{2021}]{Polzin21}
{Polzin} A.,  {van Dokkum} P.,  {Danieli} S.,  {Greco} J.~P.,   {Romanowsky}
  A.~J.,  2021, \apjl, 914, L23

\bibitem[\protect\citeauthoryear{{Robotham} et~al.,}{{Robotham}
  et~al.}{2010}]{Robotham10}
{Robotham} A.,  et~al., 2010, \pasa, 27, 76

\bibitem[\protect\citeauthoryear{{Robotham} et~al.,}{{Robotham}
  et~al.}{2011}]{Robotham11}
{Robotham} A.~S.~G.,  et~al., 2011, \mnras, 416, 2640

\bibitem[\protect\citeauthoryear{{Robotham} et~al.,}{{Robotham}
  et~al.}{2012}]{Robotham12}
{Robotham} A.~S.~G.,  et~al., 2012, \mnras, 424, 1448

\bibitem[\protect\citeauthoryear{{Robotham} et~al.,}{{Robotham}
  et~al.}{2013}]{Robotham13}
{Robotham} A.~S.~G.,  et~al., 2013, \mnras, 431, 167

\bibitem[\protect\citeauthoryear{{Robotham} et~al.,}{{Robotham}
  et~al.}{2014}]{Robotham14}
{Robotham} A.~S.~G.,  et~al., 2014, \mnras, 444, 3986

\bibitem[\protect\citeauthoryear{{Salim}}{{Salim}}{2014}]{Salim14a}
{Salim} S.,  2014, Serbian Astronomical Journal, 189, 1

\bibitem[\protect\citeauthoryear{{Salim}, {Lee}, {Ly}, {Brinchmann},
  {Dav{\'e}}, {Dickinson}, {Salzer}  \& {Charlot}}{{Salim}
  et~al.}{2014}]{Salim14}
{Salim} S.,  {Lee} J.~C.,  {Ly} C.,  {Brinchmann} J.,  {Dav{\'e}} R.,
  {Dickinson} M.,  {Salzer} J.~J.,   {Charlot} S.,  2014, preprint

\bibitem[\protect\citeauthoryear{{Sawala} et~al.,}{{Sawala}
  et~al.}{2016}]{Sawala16}
{Sawala} T.,  et~al., 2016, \mnras, 457, 1931

\bibitem[\protect\citeauthoryear{{Simon}}{{Simon}}{2019}]{Simon19a}
{Simon} J.~D.,  2019, \araa, 57, 375

\bibitem[\protect\citeauthoryear{{Sotillo-Ramos} et~al.,}{{Sotillo-Ramos}
  et~al.}{2021}]{Sotillo-Ramos21}
{Sotillo-Ramos} D.,  et~al., 2021, arXiv e-prints, p. arXiv:2109.12078

\bibitem[\protect\citeauthoryear{{Starkenburg}, {Oman}, {Navarro}, {Crain},
  {Fattahi}, {Frenk}, {Sawala}  \& {Schaye}}{{Starkenburg}
  et~al.}{2017}]{Starkenburg17}
{Starkenburg} E.,  {Oman} K.~A.,  {Navarro} J.~F.,  {Crain} R.~A.,  {Fattahi}
  A.,  {Frenk} C.~S.,  {Sawala} T.,   {Schaye} J.,  2017, \mnras, 465, 2212

\bibitem[\protect\citeauthoryear{{Tanaka}, {Chiba}, {Hayashi}, {Komiyama},
  {Okamoto}, {Cooper}, {Okamoto}  \& {Spitler}}{{Tanaka}
  et~al.}{2018}]{Tanaka18}
{Tanaka} M.,  {Chiba} M.,  {Hayashi} K.,  {Komiyama} Y.,  {Okamoto} T.,
  {Cooper} A.~P.,  {Okamoto} S.,   {Spitler} L.,  2018, \apj, 865, 125

\bibitem[\protect\citeauthoryear{{Tollerud}, {Bullock}, {Strigari}  \&
  {Willman}}{{Tollerud} et~al.}{2008}]{Tollerud08}
{Tollerud} E.~J.,  {Bullock} J.~S.,  {Strigari} L.~E.,   {Willman} B.,  2008,
  \apj, 688, 277

\bibitem[\protect\citeauthoryear{{Treyer} et~al.,}{{Treyer}
  et~al.}{2017}]{Treyer17a}
{Treyer} M.,  et~al., 2017, preprint

\bibitem[\protect\citeauthoryear{{Wang} \& {White}}{{Wang} \&
  {White}}{2012}]{Wang12c}
{Wang} W.,  {White} S. D.~M.,  2012, \mnras, 424, 2574

\bibitem[\protect\citeauthoryear{{Wang} et~al.,}{{Wang} et~al.}{2021}]{Wang21i}
{Wang} W.,  et~al., 2021, \mnras, 500, 3776

\bibitem[\protect\citeauthoryear{{Weerasooriya}, {Bovill}, {Benson}, {Musick}
  \& {Ricotti}}{{Weerasooriya} et~al.}{2023}]{Weerasooriya23}
{Weerasooriya} S.,  {Bovill} M.~S.,  {Benson} A.,  {Musick} A.~M.,   {Ricotti}
  M.,  2023, \apj, 948, 87

\bibitem[\protect\citeauthoryear{{Weinmann}, {van den Bosch}, {Yang}  \&
  {Mo}}{{Weinmann} et~al.}{2006}]{Weinmann06}
{Weinmann} S.~M.,  {van den Bosch} F.~C.,  {Yang} X.,   {Mo} H.~J.,  2006,
  \mnras, 366, 2

\bibitem[\protect\citeauthoryear{{Weisz} et~al.,}{{Weisz}
  et~al.}{2011}]{Weisz11c}
{Weisz} D.~R.,  et~al., 2011, \apj, 743, 8

\bibitem[\protect\citeauthoryear{{Wetzel}, {Hopkins}, {Kim}, {Faucher-Giguere},
  {Keres}  \& {Quataert}}{{Wetzel} et~al.}{2016}]{Wetzel16}
{Wetzel} A.~R.,  {Hopkins} P.~F.,  {Kim} J.-h.,  {Faucher-Giguere} C.-A.,
  {Keres} D.,   {Quataert} E.,  2016, preprint

\bibitem[\protect\citeauthoryear{{Wright} et~al.,}{{Wright}
  et~al.}{2016}]{Wright16}
{Wright} A.~H.,  et~al., 2016, \mnras, 460, 765

\bibitem[\protect\citeauthoryear{{Wu} et~al.,}{{Wu} et~al.}{2022}]{Wu22d}
{Wu} J.~F.,  et~al., 2022, \apj, 927, 121

\bibitem[\protect\citeauthoryear{{Zibetti}, {Charlot}  \& {Rix}}{{Zibetti}
  et~al.}{2009}]{Zibetti09a}
{Zibetti} S.,  {Charlot} S.,   {Rix} H.,  2009, \mnras, pp 1410--+

\bibitem[\protect\citeauthoryear{{da Cunha}, {Charlot}  \& {Elbaz}}{{da Cunha}
  et~al.}{2008}]{da-Cunha08}
{da Cunha} E.,  {Charlot} S.,   {Elbaz} D.,  2008, \mnras, 388, 1595

\bibitem[\protect\citeauthoryear{{de los Reyes}, {Kirby}, {Zhuang}, {Steidel},
  {Chen}  \& {Wheeler}}{{de los Reyes} et~al.}{2023}]{de-los-Reyes23a}
{de los Reyes} M. A.~C.,  {Kirby} E.~N.,  {Zhuang} Z.,  {Steidel} C.~C.,
  {Chen} Y.,   {Wheeler} C.,  2023, \apj, 951, 52

\makeatother
\end{thebibliography}







\bsp	
\label{lastpage}
\end{document}